\documentclass[12pt]{article}
\usepackage[T1]{fontenc}
\usepackage[utf8]{inputenc}
\usepackage{authblk}
\usepackage{booktabs}
\usepackage{setspace}

    \usepackage{amsfonts}
    \usepackage{amssymb}
    \usepackage{amsbsy}
    \usepackage{amsmath}
    \usepackage{amsthm}
    \usepackage{bbm}
    \usepackage{bm}
    \usepackage{mathtools}
    \usepackage{caption}
    \usepackage{subcaption}    
 
\newcommand{\lspace}[1]{\renewcommand{\baselinestretch}{#1} \small\normalsize}

    \newcounter{subreq}[req]
	
	\makeatletter
	\renewcommand{\p@subreq}{\thereq}
	\makeatother

\newcounter{subassumption}[assum]
	
	\makeatletter
	\renewcommand{\p@subassumption}{\theassum}
	\makeatother

	\newtheorem*{obs*}{Observation}

    \usepackage[flushleft,online,para]{threeparttable}
	\usepackage{graphics}
\usepackage[vlined,linesnumbered,ruled,resetcount]{algorithm2e}
\SetKwInOut{Initialization}{Initialization}

    \graphicspath{{Graphs/}}
\usepackage{adjustbox}
\usepackage{multirow}

    \usepackage{tikz}
    \usetikzlibrary{arrows}
    \usetikzlibrary{decorations.pathreplacing}
    \usepackage{pgfplots}
\usepackage{indentfirst}             
\usepackage{floatrow}
    \usepackage{tabularx}
        \newcolumntype{Z}{>{\centering\arraybackslash}X}
        \newcolumntype{L}{>{\raggedright\arraybackslash}X}
    \usepackage{dcolumn}
        \newcolumntype{d}[1]{D{.}{.}{#1}}
    \usepackage{rotating}                     
    \usepackage{lscape}
    \usepackage{pdflscape}   
\usepackage{authblk}

	\usepackage{csquotes}
	\usepackage[round]{natbib}
	\usepackage{url}

    \usepackage{xcolor}
        \definecolor{darkblue}{rgb}{0,0,0.4}
    \usepackage{hyperref}
        \hypersetup{
            colorlinks = true,
            linkcolor = darkblue,
            citecolor = darkblue,
            pdfborder = 0 0 0,
            pdfdisplaydoctitle = true,
            pdfhighlight = /N,
            pdfpagelayout = OneColumn,
            pdfpagemode = UseNone,
            pdfstartview = {FitH},
            pdfauthor = {{AB, CF \& JR}},
            pdftitle = {{null}},
            pdfsubject = {{}}
        }

    \usepackage[textsize=footnotesize, colorinlistoftodos, textwidth=4cm, obeyDraft]{todonotes}
    \usepackage{geometry}
    \usepackage{setspace}
    \usepackage[bottom, multiple]{footmisc}  
    \usepackage{verbatim}
    \usepackage[normalem]{ulem}     
    \usepackage{mathpazo}

    \usepackage[toc,page]{appendix}
	\usepackage{lipsum}

\begin{document}

\vspace{-.5in}

\title{Bayesian estimation of finite mixtures of Tobit models\thanks{\enspace I thank Blake McShane, Andrey Simonov, Adam Smith, and Leo Trigo for helpful comments, and Pankhuri Saxena for excellent research assistance. All errors are my own. E-mail address for correspondence: caio.waisman@kellogg.northwestern.edu.}}

\author{Caio Waisman\\
Kellogg School of Management\\
Northwestern University \\
}

\date{\today}

\maketitle
\thispagestyle{empty}

\lspace{1}

\begin{abstract}

\noindent This paper outlines a Bayesian approach to estimate finite mixtures of Tobit models. The method consists of an MCMC approach that combines Gibbs sampling with data augmentation and is simple to implement. I show through simulations that the flexibility provided by this method is especially helpful when censoring is not negligible. In addition, I demonstrate the broad utility of this methodology with applications to a job training program, labor supply, and demand for medical care. I find that this approach allows for non-trivial additional flexibility that can alter results considerably and beyond improving model fit. 

\vspace{.2in}
\noindent


\end{abstract}

\newpage
\setcounter{page}{1}
\lspace{1.1}

\section{Introduction}\label{sec:intro}

Consider the censored regression model from \cite{tobin1958}
\begin{align}\label{eq:tob_model}
\begin{split}
    y_i^* &\sim \text{N} \left ( x_i'\beta, \sigma^2 \right ) \\
    y_i &=   \max \{0, y_i^* \}
\end{split}
\end{align}
where $i$ denotes an observation; $x_i$ is a $p$-dimensional vector of covariates; $\beta$ is a $p$-dimensional vector of coefficients; $\sigma^2$ is the variance; $\text{N} \left (\mu, \sigma^2 \right )$ denotes the normal distribution with mean $\mu$ and variance $\sigma^2$; and where $x_i$ and $y_i$ are observed but $y_i^*$ is not.

This model, also known as the Tobit model, became popular because it enables the study of relationships between a censored dependent variable and explanatory variables. The simplicity of this model yields a well-behaved likelihood function, allowing for an easily obtainable maximum likelihood estimator (MLE)---see, for instance, \cite{amemiya1973} and \cite{olsen1978}. In addition, \cite{chib1992} demonstrated that it is also easy to perform Bayesian inference on the parameters of this model by using a simple procedure that combines Gibbs sampling with data augmentation.

Unsurprisingly, such simplicity does not come without a cost. The Tobit model imposes a very restrictive structure on the data, making it insufficiently flexible to accommodate empirical patterns that are typically observed in practice, such as skewness. This shortcoming motivated a large literature that attempts to model the relationship of interest in a more flexible way while preserving the convenience and tractability of the Tobit model as much as possible. In this paper, I propose combining a finite mixture of normal distributions as the data generating process with MCMC estimation techniques to accomplish this task.

Finite mixtures of normal distributions have several properties that make them an attractive modeling framework. For instance, they can achieve great flexibility with only a few components. In addition, there are theoretical guarantees that, given enough components, mixtures of normal distributions can approximate any density function arbitrarily well (e.g., \citealt{norets2010}). Through a series of simulations, I verify that these properties indeed hold. Interestingly, I also find that overparametrization relative to true data generating processes might be desirable to better fit the observed data when censoring is considerable.\footnote{The code used in this paper is available upon and will be made publicly available in the future.}















I further showcase the broad utility applicability of the proposed methodology via three empirical applications. Each application illustrates a different practical implication from utilizing the proposed methodology instead of the standard Tobit model. 


First, I revisit the data from \cite{lalonde1986} and \cite{dw1999} and consider the estimation of the effect of a job training program on workers' earnings. Given that the outcome variable is measured in monetary amounts and censored, the Tobit model lends itself naturally to this analysis. I find that the standard Tobit model yields the counterintuitive result that taking the program decreases earnings, whereas the mixture of normal distributions implies the opposite result. This shows how the added flexibility can lead to different conclusions regarding estimates of specific relationships between outcomes and covariates.

Second, I utilize data from the Panel Study of Income Dynamics (PSID) to study labor supply differences between married an unmarried women. The results indicate that not only does the mixture model provide a better overall fit of the data relative to the standard Tobit model, but it also exacerbates differences in the probability of not working and on the average annual hours worked between married and unmarried women. The results further demonstrate that the impacts of using the proposed model go beyond simply improving model fit.

Finally, I assess whether the proposed model is able to provide a good fit for data to which it might not be considered ideal. To this end, I apply the proposed approach to a data set that has number of visits to doctors as a measure of demand for medical care and as the outcome variable following the specification from \cite{dt1997}. Although this is a count variable, I find that the presented methodology yields a good fit despite arguably being better suited to handle continuous variables and that it is very similar, if not better, than the fit of models designed specifically to handle count outcome variables.

The model here considered has been applied in the context of real-time bidding (RTB) auctions by \cite{czlm2011} to estimate the distribution of winning bids. Despite being order statistics, the specific form of the distribution of winning bids is unknown because so are the number of bidders and the parent distribution in this setting. This renders the agnostic and flexible methodology addressed in this paper an attractive alternative to estimate this distribution. This might also be the case in other marketing settings where the standard Tobit model was used, such as salesforce compensation (\citealt{cn1992,mcn2005}), advertising spending (\citealt{hm2003, lmmr2020}), price negotiations (\citealt{biz2017, jindal2022}), and customer spending (\citealt{vlb2018, kk2018, jsct2020}). Indeed, my findings indicate that using the more flexible mixtures can change results both quantitatively and qualitatively.

The rest of the paper proceeds as follows. First, I provide a brief discussion about the related literature. Section \ref{sec:model} introduces the model and Section \ref{sec:bayes} presents the proposed Gibbs sampling approach to estimate it. Section \ref{sec:simul} provides a simulation exercise to demonstrate the validity of the procedure, whereas Section \ref{sec:applies} features applications to a real-world data sets on a job training program experiment, labor supply, and demand for medical care. Finally, Section \ref{sec:conc} concludes.

\subsection{Related Literature}\label{sec:rel_lit}

The approach this paper introduces is related to several existing methods to estimate censored regression models. Various studies proposed replacing normality with a different distributional assumption that is not as restrictive. One approach that proved itself attractive was to use mixtures of distributions because of their flexibility. In turn, two estimation and inference strategies were typically combined with these models: MLE, often through the use of EM algorithms, and Bayesian MCMC methods.

\cite{avcgfmg2012} and \cite{mcml2015} relaxed the normality assumption by instead assuming that the latent variable from equation (\ref{eq:tob_model}), $y^*$, followed a \textit{t} distribution. They then leveraged the relationship between this distribution and scale mixture of normal distributions to propose EM algorithms for estimation, whereas \cite{gblc2015} proposed an MCMC method instead.

Analogously, instead of normality \cite{hs2011} assumed that $y^*$ followed a skew-normal distribution and used maximum likelihood estimation. This approach was then extended by \cite{mgcl2017} and \cite{mgl2018}, who used scale mixtures of skew-normal distributions and proposed an MCMC method and an EM algorithm for estimation, respectively. In turn, \cite{lcpd2019} proposed using a mixture of \textit{t} distributions, while \cite{zclb2019} suggested using a mixture of scale mixtures of normal distributions. Both papers developed EM algorithms to compute the MLE for these models.

My approach differs from those because I propose modeling $y^*$ as following a finite mixture of normal distributions, which implies a finite mixture of Tobit models for the observed outcome variable, $y$. This model was perhaps first considered by \cite{jrd1993}, who suggested maximum likelihood estimation using the EM algorithm. I complement this study by demonstrating how the same estimation task can be accomplished through Bayesian methods. This is an attractive approach given that MCMC methods are often better suited to handle settings with finite mixtures of distributions than MLE approaches. (For a more detailed discussion, see, for example, Chapter 1 of \citealt{rossi2014}.)

Finally, my approach also relates to the smooth mixture of Tobit models from \cite{ks2011}. Their model is arguably richer than the mixture of Tobit models because they assume that the class an observation belongs to depends on this observation's covariates, and use a special case of a multinomial probit, where the variances of all error terms equal one, to determine this class, which aids in identification.

The richness added by \cite{ks2011} requires them to use a more involved estimation approach that consists of Gibbs sampling with a Metropolis within Gibbs step. This would be necessary even if the only covariate used in their multinomial probit was a constant, which corresponds to the mixture of Tobit models I consider. On the other hand, I model the latent class identifier as following a categorical distribution. This allows me to rely solely on Gibbs sampling with data augmentation, yielding a simpler algorithm to obtain draws from the posterior distribution of parameters given the data. Hence, I view my method as complementary to theirs. 

\section{Model}\label{sec:model}

Following \cite{jrd1993}, I replace the model in (\ref{eq:tob_model}) with:
\begin{align}\label{eq:tob_model2}
\begin{split}
    &y_i^* \sim \text{N} \left ( x_i'\beta_c, \sigma_c^2 \right ) \text{ with probability } \pi_c,\text{ where }c=1,\dots,C \\
    &\pi_c\in(0,1)\text{ for all }c\text{ and }\sum_{c=1}^C\pi_c=1 \\
    & y_i =  \min \left \{ \max \{l_i, y_i^* \}, u_i \right \}
\end{split}
\end{align}
where the censoring bounds, $l_i$ and $u_i$, are known.

In this model, the latent variable $y^*$ follows a mixture of $C$ normal distributions, which are indexed by $c$. The objective is to estimate and perform  inference on the parameters $\theta \equiv \left \{\beta_c, \sigma_c^2, \pi_c \right \}_{c=1}^C$. Hence, there are $(2+p)\times C-1$ parameters to be estimated.

Relative to \cite{jrd1993}, this model also relaxes two features of the typical censored regression model used in numerous studies: it allows for two-sided censoring, that is, for the observed outcome variable, $y$, to be censored from below at $l$ or from above at $u$; and it further allows these censoring levels to be observation-specific. For the purposes of outlining the Gibbs sampling algorithm given in Section \ref{sec:bayes}, allowing for observation-specific two-sided censoring is straightforward. However, in most applications these bounds are the same across all observations.

    
\section{Bayesian Inference: A Gibbs Sampling Approach}\label{sec:bayes}

This section outlines the MCMC method to estimate the parameters of the model given in (\ref{eq:tob_model2}). It is a simple Gibbs sampling procedure with data augmentation that combines the approach used to estimate the standard Tobit model introduced by \cite{chib1992} with the typical method to estimate the parameters of finite mixtures of normal distributions (e.g., \citealt{dr1994}).

The presentation features three steps. First, I present the likelihood function of this model, followed by the prior distributions used for estimation. Then, I introduce the Gibbs sampling algorithm to obtain draws from the posterior distribution of the parameters given the observed data.

\subsection{Likelihood Function}\label{sec:likel}

The data consist of $n$ i.i.d. observations indexed by $i$: $\left \{y_i,x_i,l_i,u_i. \right \}_{i=1}^n$. Conditional on belonging to class $c$, the likelihood function for observation $i$ is
\begin{align}\label{eq:like_i_cond}
    \ell_{ic}(\theta)= \left [\Phi \left (\frac{l_i-x_i'\beta_c}{\sigma_c} \right ) \right ]^{\mathbbm{1}\{y_i=l_i\}} \left [\frac{1}{\sigma_c} \phi \left (\frac{y_i-x_i'\beta_c}{\sigma_c} \right ) \right ]^{\mathbbm{1}\{l_i<y_i<u_i\}}\left [\Phi \left (\frac{x_i'\beta_c-u_i}{\sigma_c} \right ) \right ]^{\mathbbm{1}\{y_i=u_i\}}
\end{align}
where $\phi(\cdot)$ and $\Phi(\cdot)$ denote the pdf and the cdf of the standard normal distribution, respectively.

Integrating with respect to the classes, the likelihood for observation $i$ becomes
\begin{align}\label{eq:like_i}
    \ell_{i}(\theta)= \sum_{c=1}^C \pi_c \ell_{ic}(\theta).
\end{align}
Because the observations are i.i.d, the likelihood of the data is
\begin{align}\label{eq:like}
    \ell(\theta)=\prod_{i=1}^n \ell_{i}(\theta)
\end{align}
and so the log-likelihood is
\begin{align}\label{eq:log_like}
    \log\ell(\theta)=\sum_{i=1}^n \log\ell_{i}(\theta)=\sum_{i=1}^n \log \left ( \sum_{c=1}^C \pi_c \ell_{ic}(\theta)\right ).
\end{align}

\subsection{Prior Distributions}\label{sec:prior}

To implement the algorithm and obtain draws from the posterior distribution of $\theta$ given the data, it is necessary to choose priors for these parameters. I choose the following prior distributions
\begin{align}\label{eq:prior}
    \begin{split}
        \pi&\sim \text{Dirichlet}\left (\alpha_1,\dots,\alpha_C \right )\\
        \begin{split}
        \beta_c &\sim \text{N} \left (\mu_c,\Omega_c^{-1} \right )\\
        \sigma_c^2 &\sim \text{IG} \left (a_c,b_c \right )
    	\end{split}, \quad c=1,\dots, C
    \end{split}
\end{align}
where: $\pi\equiv \left [\pi_1,\dots,\pi_C \right ]'$; the $\alpha$s are positive scalars; the $\mu$s are $p$-by-1 vectors; the $\Omega$s are $p$-by-$p$ matrices; and $\text{IG}(a,b)$ denotes the inverse-gamma distribution with shape parameter $a$ and scale parameter $b$.\footnote{I follow the convention where if $X\sim \text{IG} \left (a,b \right )$, then the density of $X$ is $f(x)=\frac{b^a}{\Gamma(a)}x^{-a-1}\exp\left\{-\frac{b}{x}\right\}$, where $\Gamma(\cdot)$ is the gamma function. } 
As it will become clear below, I use because they ease the implementation of the algorithm. 

\subsection{Gibbs Sampling Algorithm}\label{sec:gibbs}

Because of the censoring and the unobserved class to which a unit belongs, it is not convenient to directly work with the posterior distribution of $\theta$ given the data. However, it is straightforward to use Gibbs sampling with data augmentation to obtain draws from this distribution. I now describe this procedure, which combines the usual methods to estimate censored regression models with those used to estimate finite mixtures of normal distributions. 

\subsubsection{Prior Distribution on Class Identifiers}\label{sec:prior_z}

Following methods to estimate the parameters of finite mixtures of normal distributions, I augment the data with a latent class identifier. Denote this identifier by $z_i$, so that $z_i\in\{1,\dots,C\}$. For the sake of convenience, I assume the following prior distribution
 \begin{align}\label{eq:prior_z}
    \begin{split}
        z_i|\pi&\overset{\text{i.i.d.}}\sim \text{Categorical}\left (\pi_1,\dots,\pi_C \right )
    \end{split}
\end{align}
Given a value of $\pi$, it is straightforward to draw a random identifier $z_i$. Notice that doing so is redundant when $C=1$ and hence this step should be skipped in this case, rendering the algorithm identical to that of \cite{chib1992}. 

\subsubsection{Drawing Missing Outcomes and Augmentation}\label{sec:y_aug}

The next step in the procedure is to draw the missing values of the outcome variable due to censoring. This step is similar to the one introduced in \cite{chib1992} with a few differences, the most relevant of which is that the distributions from which the missing values are drawn are now class-specific, and thus depend on $z_i$. 

More concretely, denote the missing values of the outcome variable by $y_i^m$. Conditional on values for $\theta$, the latent class identifiers, $z_i$, and the observed data, it follows that
\begin{align}\label{eq:ymiss}
	y_i^m | y_i, x_i, u_i, l_i, z_{i}=c ,  \theta  \sim \begin{cases} \text{TN} (x_i'\beta_c,\sigma_c^2, -\infty, l_i) \text{ if } y_i=l_i \\ \text{TN} (x_i'\beta_c,\sigma_c^2, u_i, +\infty) \text{ if } y_i=u_i \end{cases}
\end{align}
where $\text{TN}(\beta,\sigma^2,l,u)$ denotes the normal distribution with mean $\beta$, variance $\sigma^2$, and truncated from below at $l$ and from above at $u$.

Having obtained draws of these missing values, the augmented outcome variable, $\tilde{y}_i$, is given by 
\begin{align}\label{eq:ytil}
	\tilde{y}_i=\mathbbm{1}\{l_i<y_i<u_i\}\times y_i+\mathbbm{1}\{y_i=l_i\text{ or } y_i=u_i\} \times y_i^m.
\end{align}

\subsubsection{Drawing Parameters of Normal Distributions}\label{sec:post_theta}

Using the augmented outcome variable, it is trivial to draw new values for the $\beta$s and the $\sigma^2$s. To do so, I first establish some new notation. Define the $n$-by-1 vector $z$, which stacks the values of $z_i$ across the $n$ observations. Further define $n_c \equiv \sum_{i=1}^n \mathbbm{1}\{z_{i}=c\}$. Let $\tilde{y}_c$ be an $n_c$-by-1 vector that stacks the values of $\tilde{y}_i$ of the observations with $z_{i}=c$. Further define the $n_c$-by-$p$ matrix $X_c$, which stacks the rows $x_i'$, analogously. It follows that
 \begin{align}\label{eq:post_theta}
    \begin{split}
        \sigma_c^2&|\beta_c,\tilde{y}_c,X_c,z,a_c,b_c\sim  \text{IG} \left (a_c + \frac{n_c}{2},b_c + \frac{(\tilde{y}_c-X_c\beta_c)'(\tilde{y}_c-X_c\beta_c)}{2} \right ) \\
	\beta_c&|\sigma_c^2,\tilde{y}_c,X_c,z,\mu_c,\Omega_c\sim \text{N} \left ( \left [ \frac{X_c'X_c}{\sigma_c^2} + \Omega_c \right ]^{-1} \left [ \frac{X_c'y_c}{\sigma_c^2} + \Omega_c \mu_c \right ], \left [ \frac{X_c'X_c}{\sigma_c^2} + \Omega_c \right ]^{-1} \right )
    \end{split}
\end{align}

\subsubsection{Drawing New Class Identifiers}\label{sec:post_z}

Next, the algorithm draws new values of the class identifiers, $z_i$. Doing so is trivial because
 \begin{align}\label{eq:post_z}
    \begin{split}
        z_i|\theta,\tilde{y}_i,x_i&\sim \text{Categorical}\left ( \frac{\pi_1 \phi \left ( \frac{\tilde{y}_i-x_i'\beta_1}{\sigma_1} \right )}{\sum_{c=1}^C \pi_c \phi \left ( \frac{\tilde{y}_i-x_i'\beta_c}{\sigma_c} \right )}, \cdots, \frac{\pi_C \phi \left ( \frac{\tilde{y}_i-x_i'\beta_C}{\sigma_C} \right )}{\sum_{c=1}^C \pi_c \phi \left ( \frac{\tilde{y}_i-x_i'\beta_c}{\sigma_c} \right )} \right  )
    \end{split}
\end{align}

\subsubsection{Drawing Class Probabilities}\label{sec:post_xi}

Finally, a round of iteration ends by drawing new values of the class probabilities, $\pi$. Conditional on $z$, it is straightforward to draw these values  because
 \begin{align}\label{eq:post_xi}
    \begin{split}
        \pi|z \sim \text{Dirichlet}\left (\alpha_1+n_1,\dots,\alpha_C+n_C \right ) 
    \end{split}
\end{align}

\subsubsection{Summary}

To summarize, the algorithm starts with initial values of $\theta$ and $z$. It then follows the steps from (\ref{eq:ymiss}) to (\ref{eq:post_xi}) to obtain a new draw from the posterior distribution of $\theta$ given the data. The process is iterated to obtain numerous such draws so that Bayesian inference can be performed on the parameters of interest.

\section{Simulation Exercises}\label{sec:simul}






One important requirement of this model is the ability to recover the mixture distribution when the true underlying DGP is indeed a finite mixture of normal distributions. I now conduct simulation exercises to demonstrate that this is the case.

\subsection{Setup}\label{sec:setup_simul}

I consider DGPs of the following form
\begin{align}\label{eq:simul_dgp}
\begin{split}
    &y_i^* \sim \text{N} \left ( \beta_{0c} + \beta_{1c} x_{1i}, \sigma_c^2 \right ) \text{ with probability } \pi_c,\text{ where }c=1, 2, 3 \\
    & x_{i}\sim \text{N}(0,4) \\
    &\pi_c\in(0,1)\text{ for all }c\text{ and }\sum_{c=1}^3 \pi_c=1 \\
    & y_i =  \min \left \{ \max \{0, y_i^* \}, 7.5 \right \}
\end{split}
\end{align}

In what follows, I keep the values of the $\pi$s, the $\beta_1$s, and the $\sigma$s fixed and vary the intercepts, $\beta_0$. Doing so allows me to bring the normal components closer together and also adjust the extent of censoring. These two elements are the main features of the distribution that can impact the performance of the estimator: the closer the components are and the more censored the data are, the harder it is for the method to recover the true mixture distribution. Table \ref{tab:common_pars} displays the values of the common parameters across DGPs.

\begin{table}[H]
\begin{threeparttable}
\caption{Common Parameters Across DGPs}
\begin{tabular}{c|ccc}
    \hline \hline 
 \multirow{2}{*}{Parameter} & \multicolumn{3}{c}{Component} \\ \cline{2-4}
  & 1 & 2 & 3 \\
\hline 
 $\beta_1$ & -0.1 & 0.2 & -0.4 \\
 $\sigma$ & 0.25 & 0.2  & 1 \\
 $\pi$ & 0.25 & 0.35 & 0.4 \\
 \hline \hline
    \end{tabular}
  \label{tab:common_pars}
  \end{threeparttable}
\end{table} 

\subsection{Data Generating Processes (DGPs)}

\begin{figure}[H]
    \begin{minipage}{\linewidth}
  \centering
  \begin{tabular}{cc}
        \includegraphics[width=0.35\textwidth]{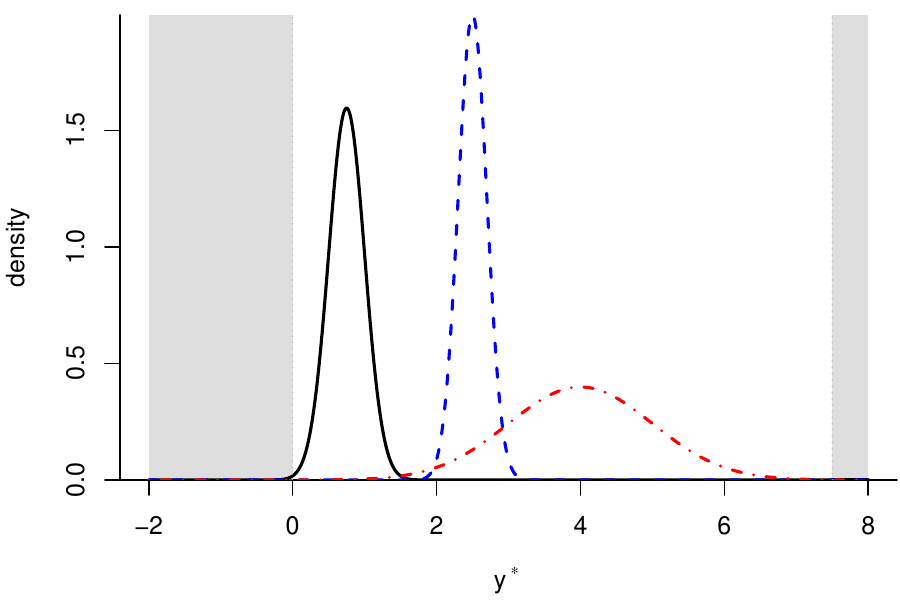} &
        \includegraphics[width=0.35\textwidth]{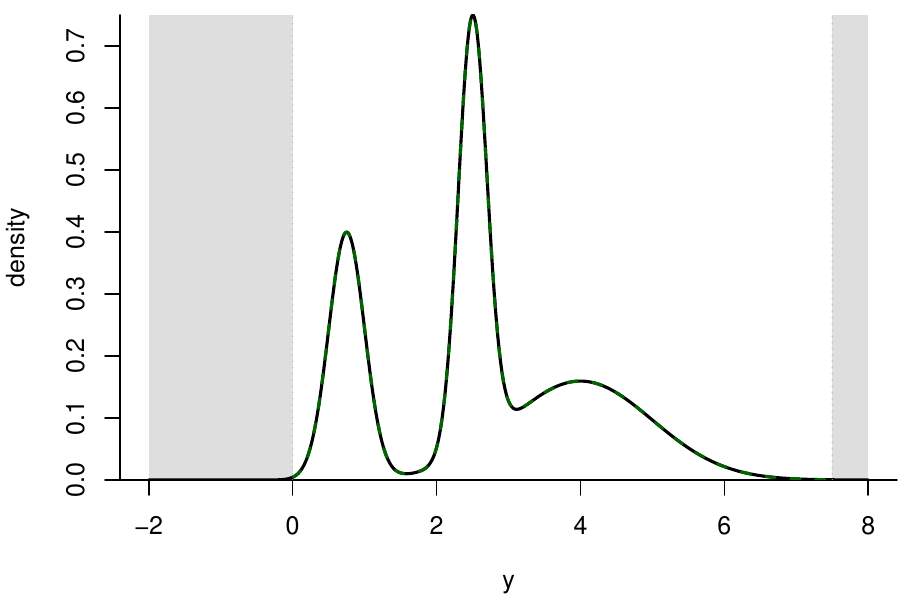} \\
        \scriptsize Components & \scriptsize Mixture
  \end{tabular}
  \subcaption{DGP 1 \label{fig:true_dists1}}
  \end{minipage}
    \begin{minipage}{\linewidth}
  \centering
  \begin{tabular}{cc}
        \includegraphics[width=0.35\textwidth]{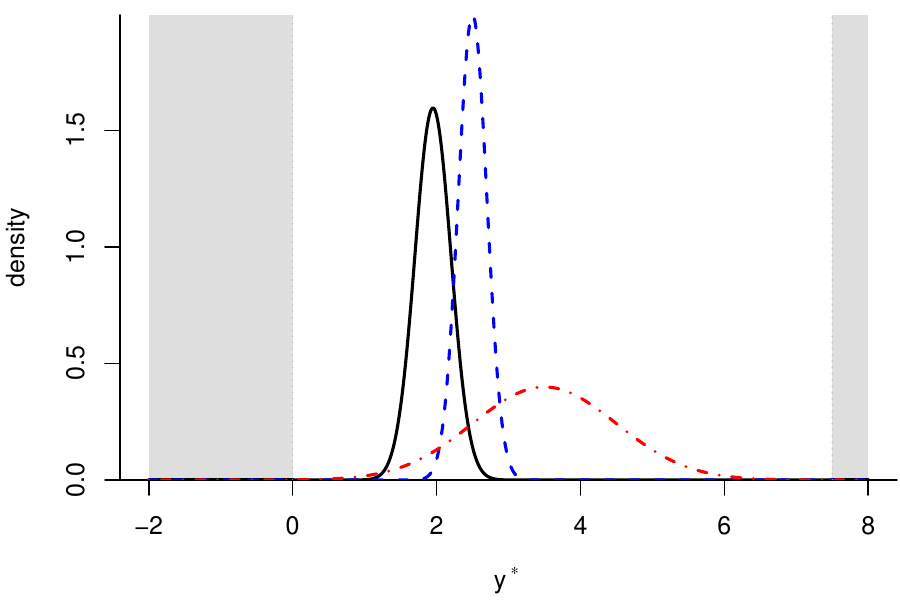} &
        \includegraphics[width=0.35\textwidth]{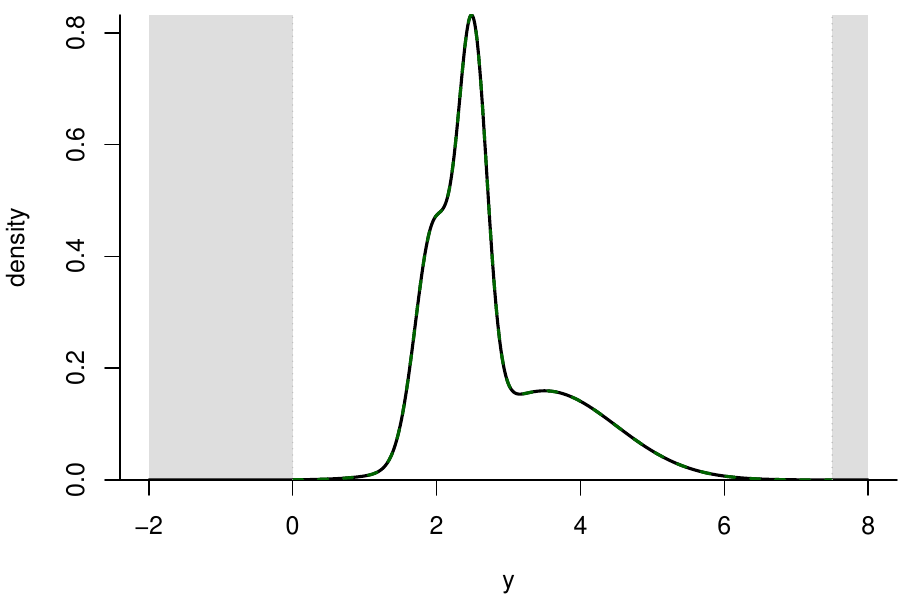} \\
        \scriptsize Components & \scriptsize Mixture
  \end{tabular}
  \subcaption{DGP 2 \label{fig:true_dists2}}
  \end{minipage}
    \begin{minipage}{\linewidth}
  \centering
  \begin{tabular}{cc}
        \includegraphics[width=0.35\textwidth]{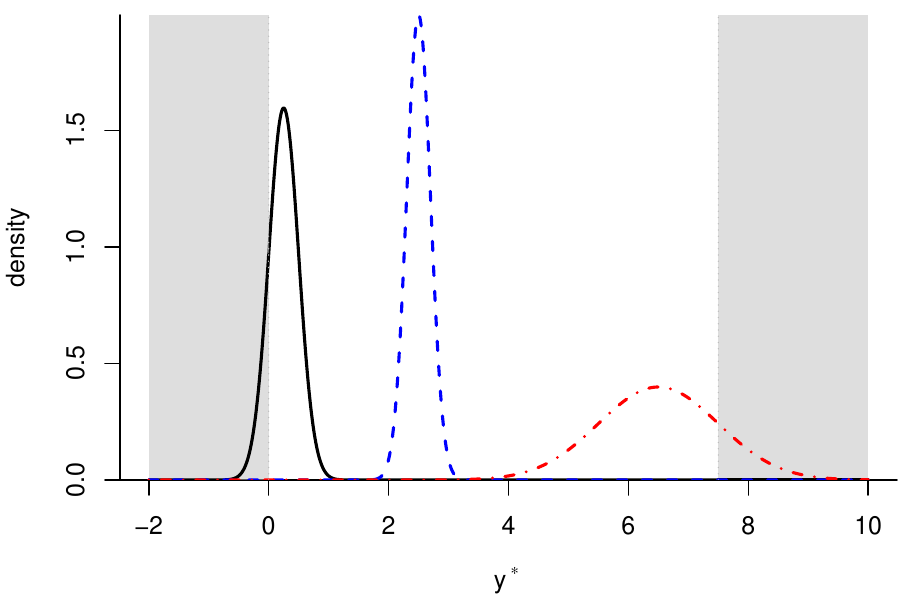} &
        \includegraphics[width=0.35\textwidth]{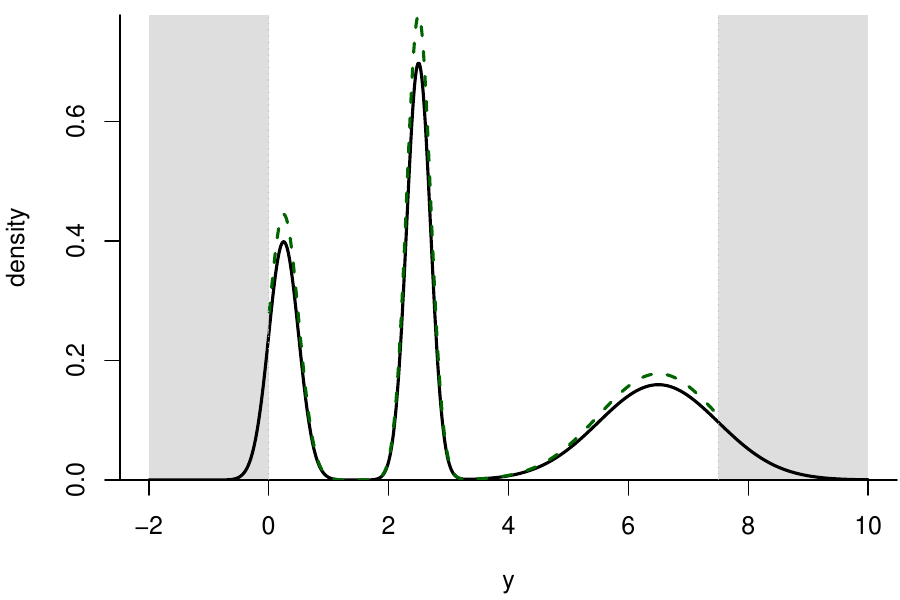} \\
        \scriptsize Components & \scriptsize Mixture
  \end{tabular}
  \subcaption{DGP 3 \label{fig:true_dists3}}
  \end{minipage}
    \begin{minipage}{\linewidth}
  \centering
  \begin{tabular}{cc}
        \includegraphics[width=0.35\textwidth]{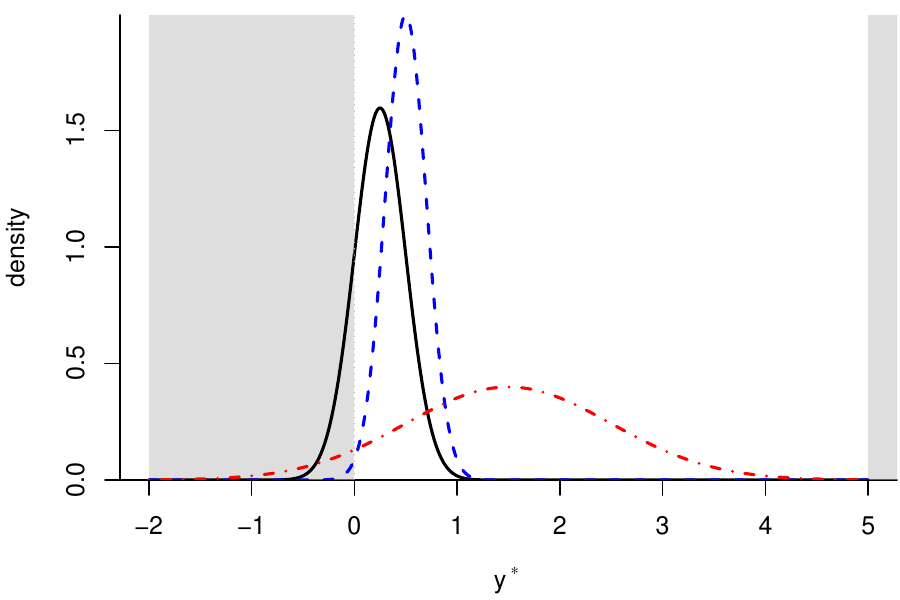} &
        \includegraphics[width=0.35\textwidth]{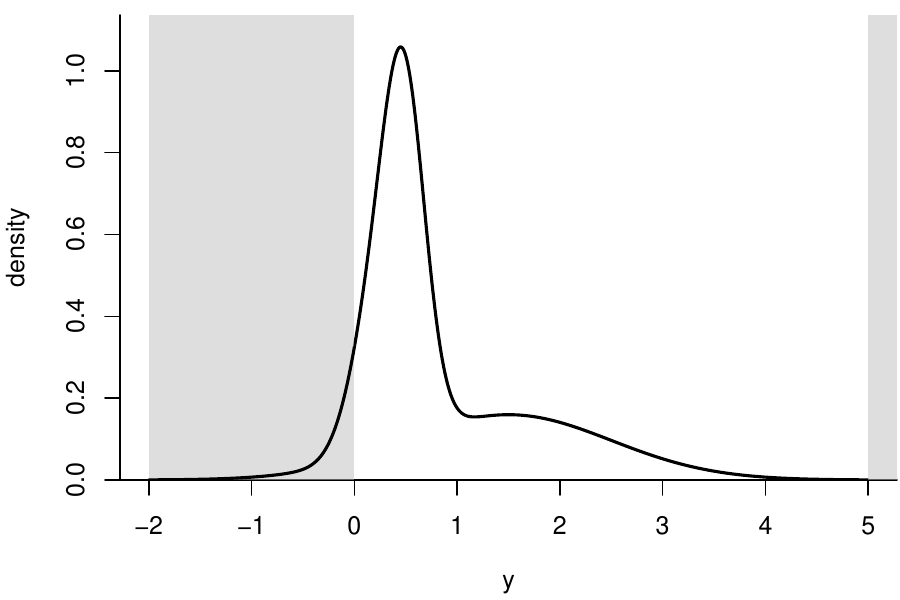} \\
        \scriptsize Components & \scriptsize Mixture
  \end{tabular}
  \subcaption{DGP 4 \label{fig:true_dists4}}
  \end{minipage}
\caption{Distributions of Components and Mixtures Across DGPs}
    \label{fig:true_dens}%
\end{figure}

I consider four DGPs, which are depicted in Figure \ref{fig:true_dens}. All plots are evaluated at $x=0$. The left column shows the three components of the mixtures, and the right column shows the resulting mixture with a solid line and the censored mixture with a dashed line. The gray regions show the areas with censoring. For completeness, the different intercepts across the DGPs are shown in Table \ref{tab:vary_pars}.

\begin{table}[H]
\begin{threeparttable}
\caption{Intercepts ($\beta_0$s) Across DGPs}
\begin{tabular}{c|ccc}
    \hline \hline 
 \multirow{2}{*}{DGP} & \multicolumn{3}{c}{Component} \\ \cline{2-4}
 & 1 & 2 & 3 \\
\hline 
 1 & 0.75 & 2.5 & 4 \\
 2 & 1.95 & 2.5 & 2.5 \\
 3 & 0.25 & 2.5 & 6.5 \\
 4 & 0.25 & 0.5 & 1.5 \\
 \hline \hline
    \end{tabular}
  \label{tab:vary_pars}
  \end{threeparttable}
\end{table} 

The first DGP, shown in Figure \ref{fig:true_dists1}, is ``well-behaved'' as the extent of censoring is minimal and there is clear separation between the normal components; consequently, the same pattern can be seen in the mixture distribution, which is then virtually indistinguishable from the censored mixture. The limited extent to which there is censoring implies that the method from Section \ref{sec:bayes} effectively becomes the typical approach to estimate finite mixtures of normal distributions as in \cite{dr1994}, for example.

The second DGP, shown in Figure \ref{fig:true_dists2}, brings the components closer together, which has ambiguous consequences for the performance of the estimator: on the one hand, bringing the distributions closer can make it more difficult for the method to detect the different components; on the other hand, it diminishes the extent of censoring. The closeness is reflected in the mixture distribution, whose modes are closer than those of the mixture distribution under DGP 1. In addition, the censoring is even less present than under DGP 1. Hence, the proposed method should perform similarly to the usual approach to estimating finite mixtures of normal distributions when these are difficult to discern.

The third DGP, displayed in Figure \ref{fig:true_dists3}, considers the opposite situation, in which the components are more distant from one another but the extent of censoring is higher. Applying the proposed estimator to data obtained under this DGP illustrates how well the method can deal with censoring when the components of the mixture are otherwise easily discernible. 

Finally, Figure \ref{fig:true_dists4} shows the fourth DGP, which, in theory, presents the most difficult setting for the estimator, in which the components are close together and censoring is more consequential. These adverse consequences are better seen in the mixture distribution, whose censoring is even more noticeable and in which only two modes are now visible.

\subsection{Simulation Results}\label{sec:res_simul}

I now present the results for each DGP-model combination. For each DGP, I create 200 data sets, each of which with 5,000 observations. I implement the algorithm described in Section \ref{sec:bayes} to each data set and fit models with one, two, three, and four components. I set: $\alpha_c=1/C$, $\mu_c=0_p$, $\Omega_c= 10 I_p$; and $a_c=b_c=0$ for all $c$. The initial value of $\pi$ to obtain the initial class identifiers was $\alpha$. I run the algorithm to obtain 2,000 draws of the parameters. I then drop the first half to handle burn-in. The starting values for the $\beta$s were $0_p$ and for the $\sigma$s were 1.

Figures \ref{fig:app1}-\ref{fig:app4} display the true DGPs with a solid line, and, across the 1,000 simulations, the average approximation with a dashed line, and the 5th and 95th percentiles around the averages with dotted lines.

\subsubsection{DGP 1}\label{sec:res_simul1}

\begin{figure}[H]
    \centering
    \begin{subfloat}[ One component \label{fig:app11}]
        {\includegraphics[width=0.49\textwidth]{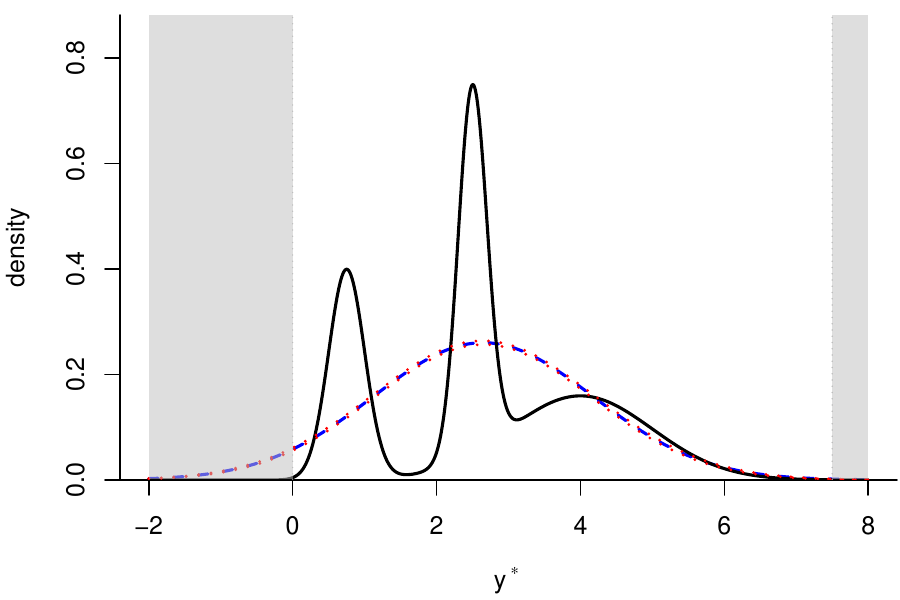}}
    \end{subfloat} 
    \begin{subfloat}[ Two components \label{fig:app12}]
        {\includegraphics[width=0.49\textwidth]{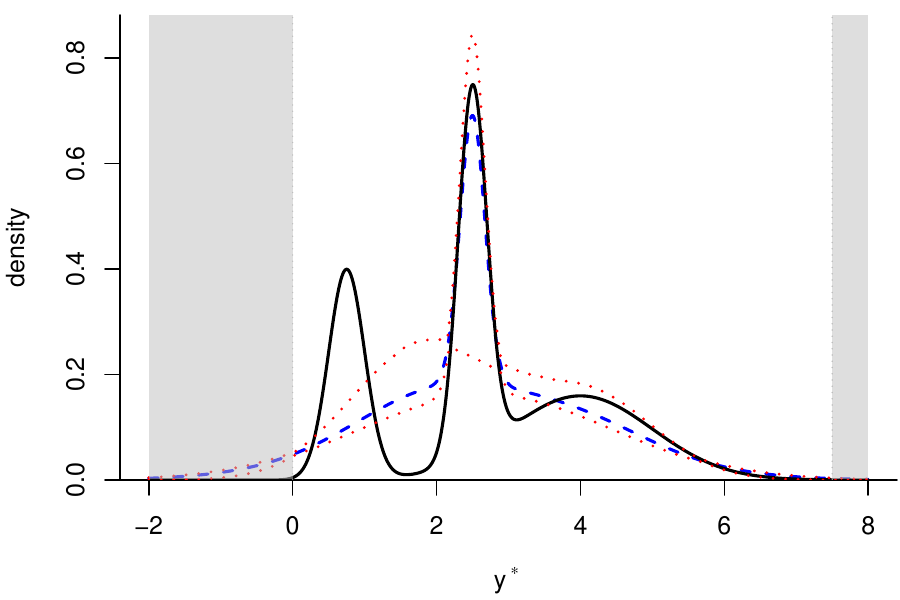}}
    \end{subfloat} \\ 
    \begin{subfloat}[ Three components \label{fig:app13}]
        {\includegraphics[width=0.49\textwidth]{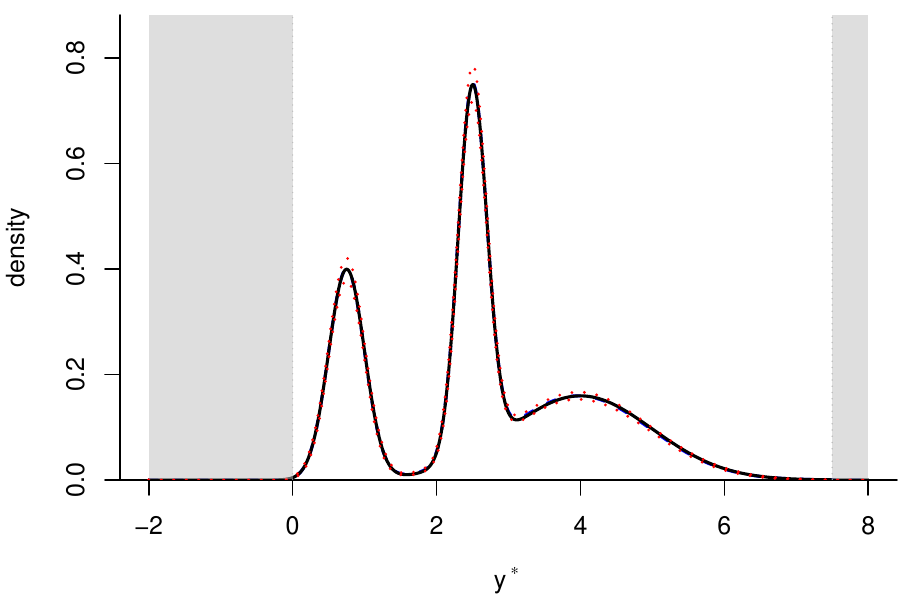}}
    \end{subfloat} 
    \begin{subfloat}[ Four components \label{fig:app14}]
        {\includegraphics[width=0.49\textwidth]{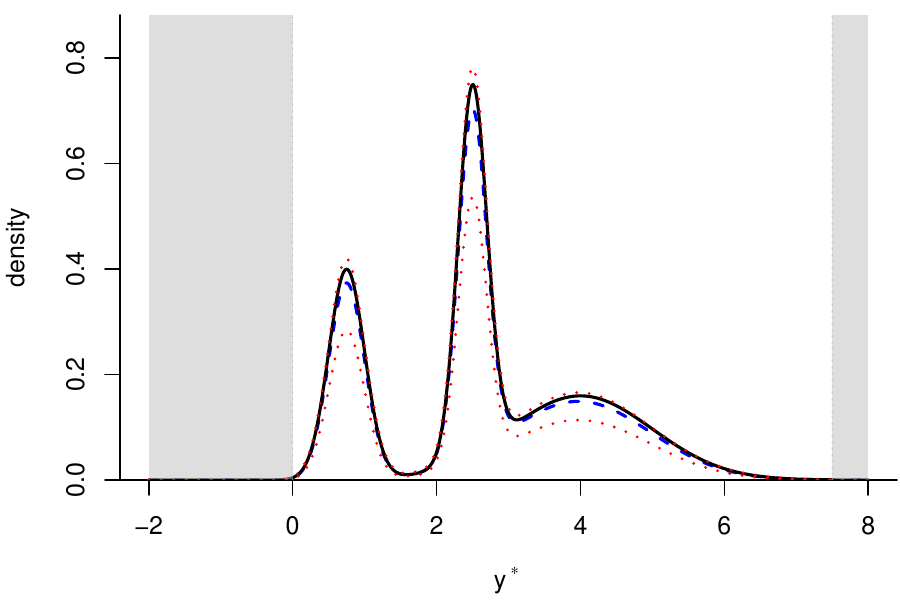}}
    \end{subfloat} 
\caption{Approximations to DGP 1}
    \label{fig:app1}%
\end{figure}

Figure \ref{fig:app11} shows the fit of a model with just one component, that is, a standard Tobit model, relative to the mixture distribution. As shown in Figure \ref{fig:true_dists1}, the DGP features three clear modes; hence, it is not surprising that this approximation differs substantially from the underlying DGP. In particular, because it cannot separate the different modes this model combines them together in a way that yields a relatively flat density function. However, it is also noteworthy that this approximation is very precisely estimated, which follows from the well-behaved nature of the standard Tobit model and the small number of parameters.

The results of adding a second component are shown in Figure \ref{fig:app12}. The model with two components better approximates the frequency associated with the distribution whose mode is relatively more common. However, this approximation is more imprecise. This is perhaps a result of idiosyncrasies across data sets that might highlight different modes more or less. As a consequence, a mixture with two components would capture different modes across samples, decreasing its precision. 

The underlying DGP is a mixture of three normal distributions. Thus, fitting this same model to the data expectedly produces good and precise approximations, as Figure \ref{fig:app13} shows. Interestingly, adding a fourth component, which overparametrizes the model relative to the underlying DGP, does not yield substantially worse approximations than the correctly specified model. Nevertheless, the approximations become less precise, as Figure \ref{fig:app14} shows.

\subsubsection{DGP 2}\label{sec:res_simul2}

\begin{figure}[H]
    \centering
    \begin{subfloat}[ One component \label{fig:app21}]
        {\includegraphics[width=0.49\textwidth]{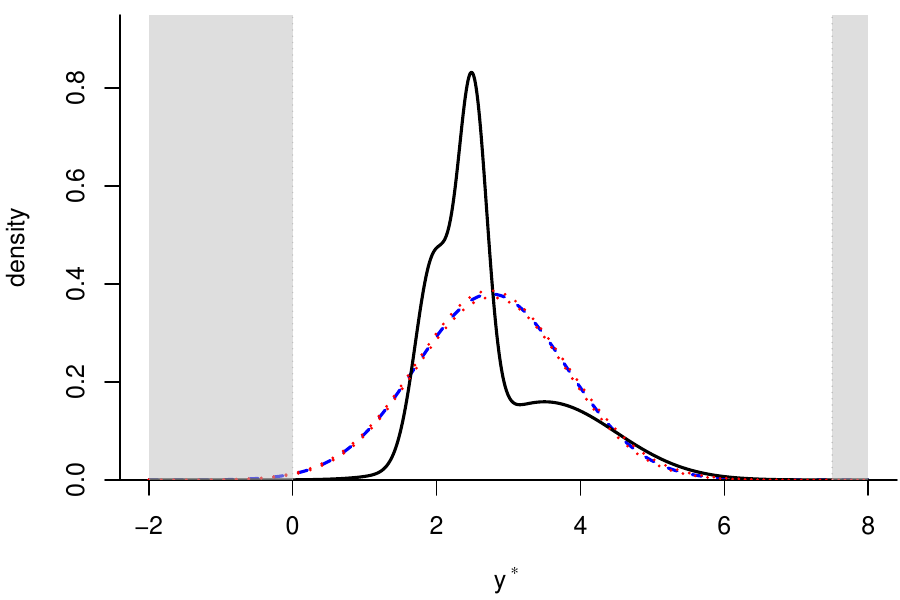}}
    \end{subfloat} 
    \begin{subfloat}[ Two components \label{fig:app22}]
        {\includegraphics[width=0.49\textwidth]{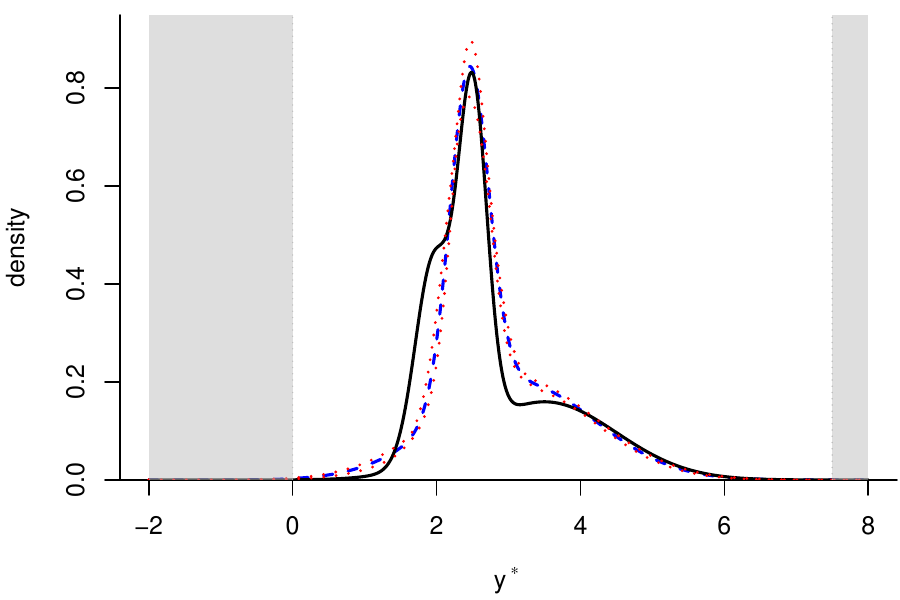}}
    \end{subfloat} \\ 
    \begin{subfloat}[ Three components \label{fig:app23}]
        {\includegraphics[width=0.49\textwidth]{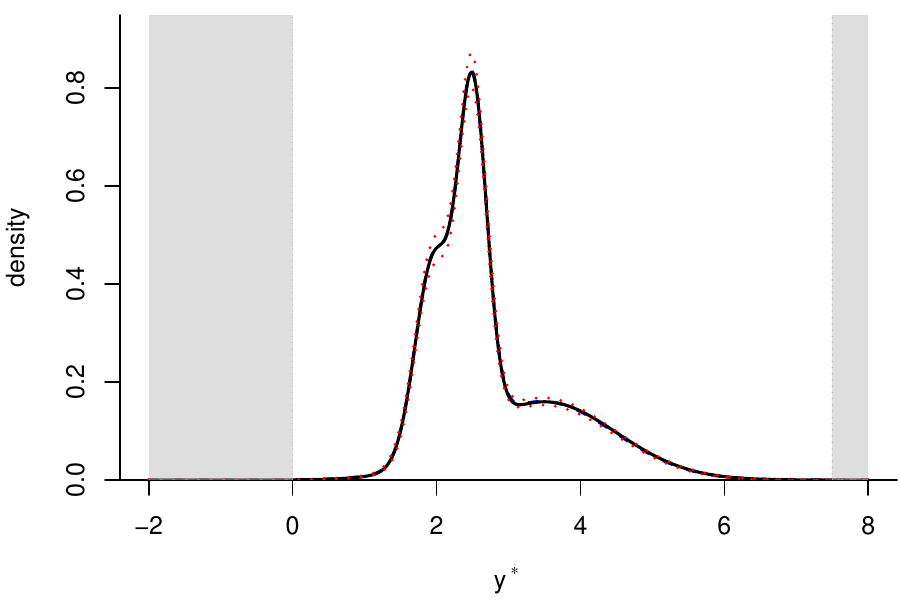}}
    \end{subfloat} 
    \begin{subfloat}[ Four components \label{fig:app24}]
        {\includegraphics[width=0.49\textwidth]{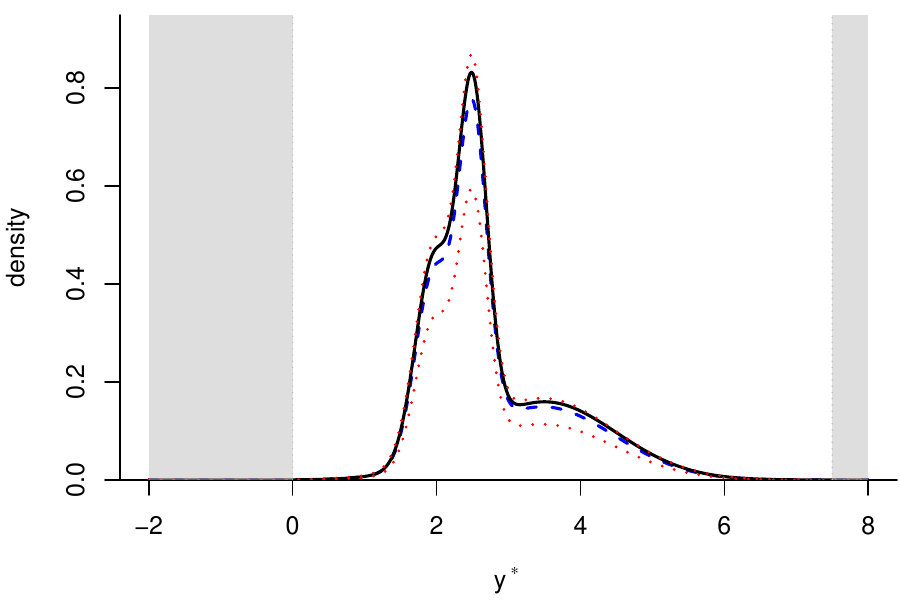}}
    \end{subfloat} 
\caption{Approximations to DGP 2}
    \label{fig:app2}%
\end{figure}

Figure \ref{fig:app2} shows results under DGP 2 in an analogous way to Figure \ref{fig:app1}, and the patterns are generally the same. All models seem to be precisely estimated, with the standard Tobit model yielding a poor approximation and the correctly specified mixture with three components matching the underlying DGP almost perfectly. Furthermore, while the model with four components still performs very similarly to the model with three components, now the model with just two components arguably performs better than when applied to DGP 1. This could be a result of the new mixture distribution featuring more concentrated modes and, consequently, in a sense being smoother, so that fewer components suffice to approximate it well.

Overall, these results reflect the ability of the usual Gibbs sampling procedure with data augmentation to estimate finite mixtures of normal distributions when the underlying components are not clearly separated.

\subsubsection{DGP 3}\label{sec:res_simul3}

\begin{figure}[H]
    \centering
    \begin{subfloat}[ One component \label{fig:app31}]
        {\includegraphics[width=0.49\textwidth]{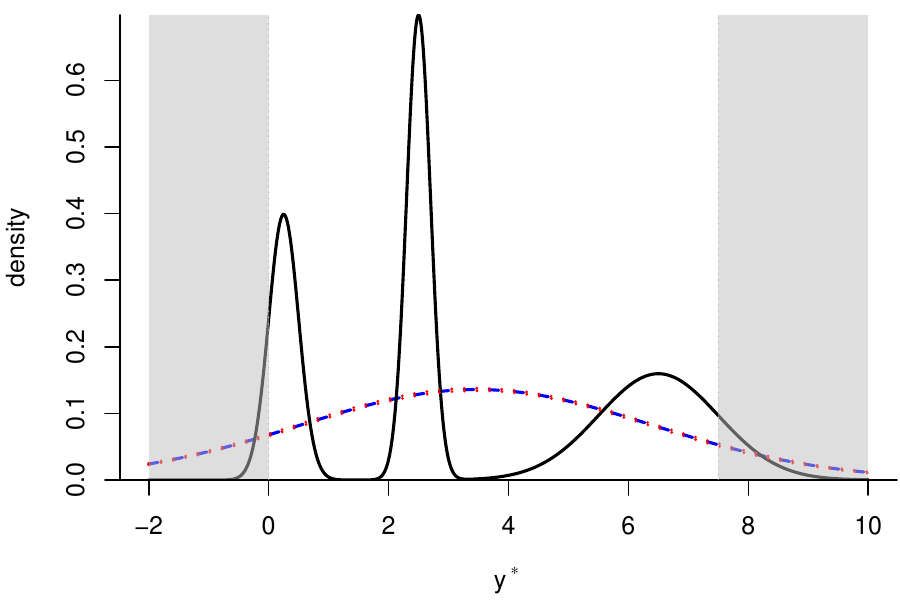}}
    \end{subfloat} 
    \begin{subfloat}[ Two components \label{fig:app32}]
        {\includegraphics[width=0.49\textwidth]{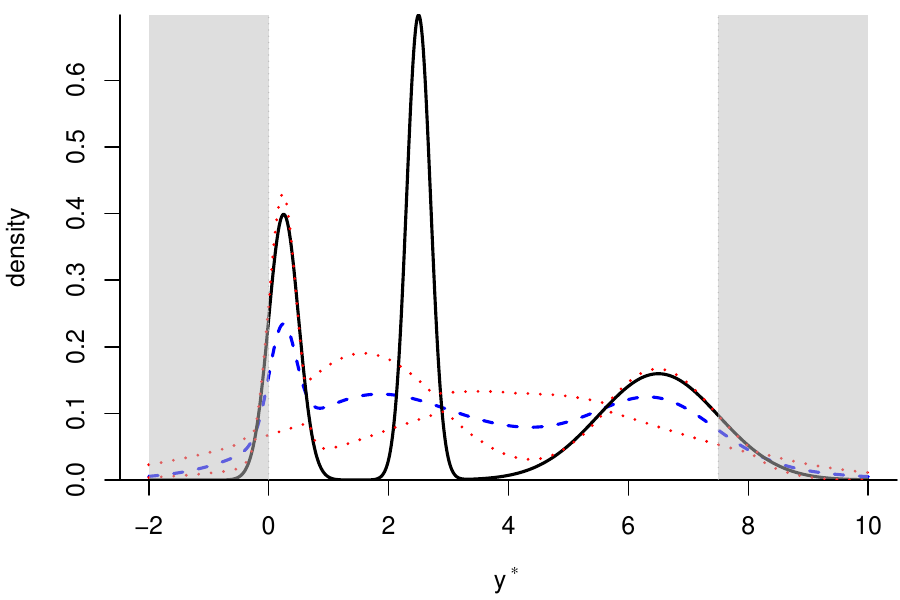}}
    \end{subfloat} \\ 
    \begin{subfloat}[ Three components \label{fig:app33}]
        {\includegraphics[width=0.49\textwidth]{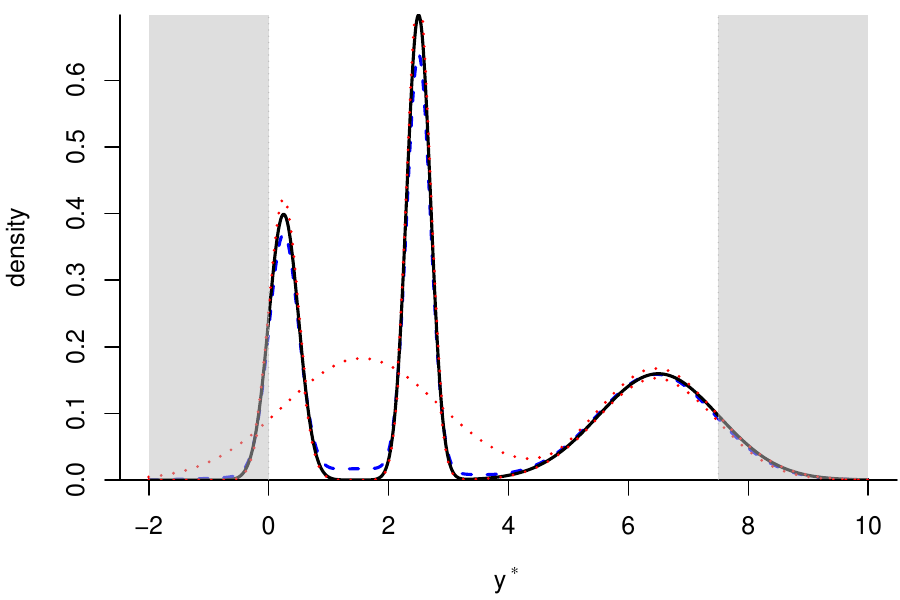}}
    \end{subfloat} 
    \begin{subfloat}[ Four components \label{fig:app34}]
        {\includegraphics[width=0.49\textwidth]{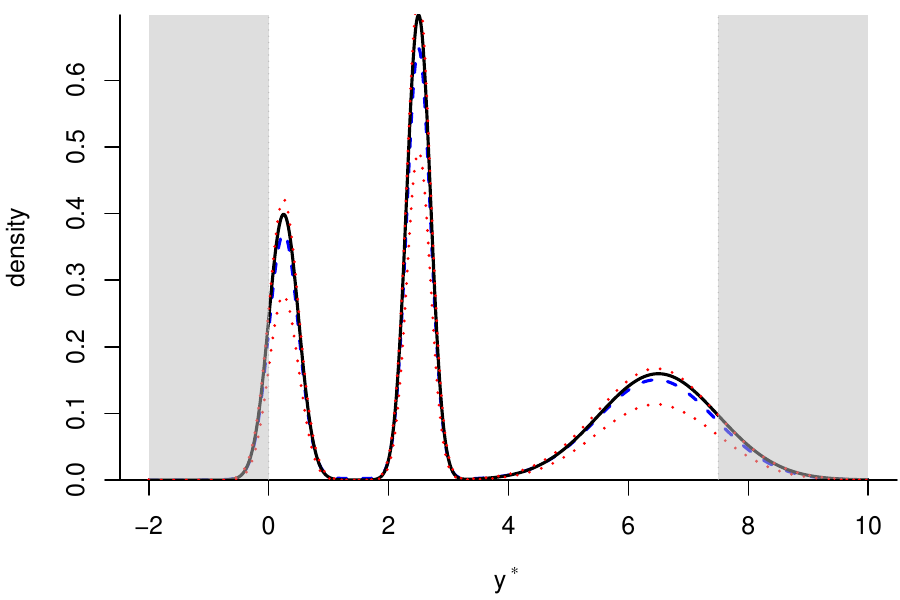}}
    \end{subfloat} 
\caption{Approximations to DGP 3}
    \label{fig:app3}%
\end{figure}

The results for the approximations of DGP 3 are displayed in Figure \ref{fig:app3}. As shown in Figure \ref{fig:true_dens}, this DGP has three well separated components, but the extent of censoring is noticeable. Thus, these results illustrate the ability of the estimator to deal with the censoring.

The approximations from using the standard Tobit model and a mixture with only two components, shown in Figures \ref{fig:app31} and \ref{fig:app32}, respectively, echo the results from DGPs 1 and 2. The former yields poor but precisely estimated approximations, while the latter provides somewhat better but more imprecise estimates. 

The results from using mixtures with three and four components show novel patterns. Figure \ref{fig:app33} displays the approximation from the model with three components, which matches the true underlying DGP. Consequently, the average approximation is good. However, the censoring now implies that this approximation yields noisy results.

In turn, using the model with four components, shown in Figures \ref{fig:app34}, not only provides a slightly better approximation than the model with just three, but the estimates are now much more precise. This suggests that having a larger number of components in the model is more attractive when the extent of censoring is more severe, even if the model becomes overparametrized relative to the true underlying DGP.

\subsubsection{DGP 4}\label{sec:res_simul4}

\begin{figure}[H]
    \centering
    \begin{subfloat}[ One component \label{fig:app41}]
        {\includegraphics[width=0.49\textwidth]{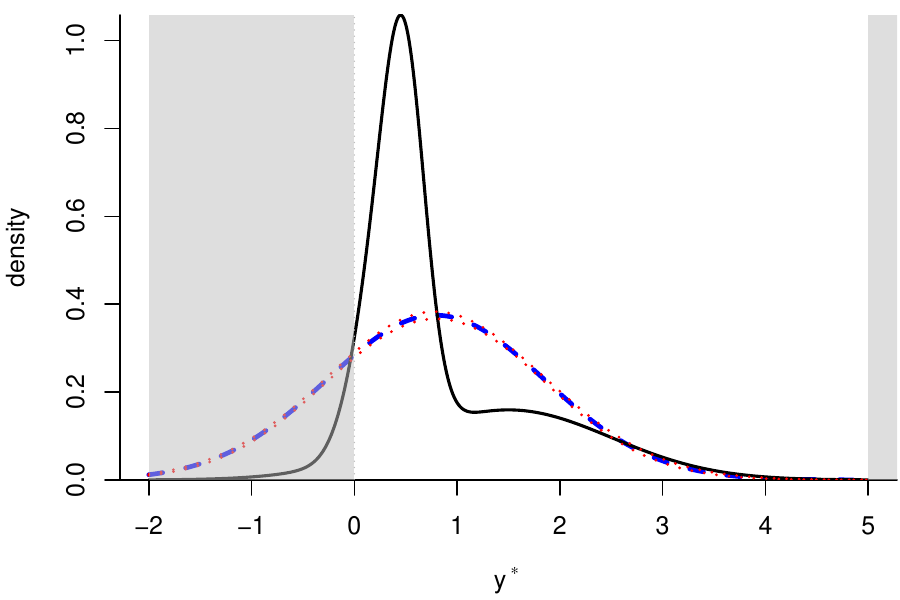}}
    \end{subfloat} 
    \begin{subfloat}[ Two components \label{fig:app42}]
        {\includegraphics[width=0.49\textwidth]{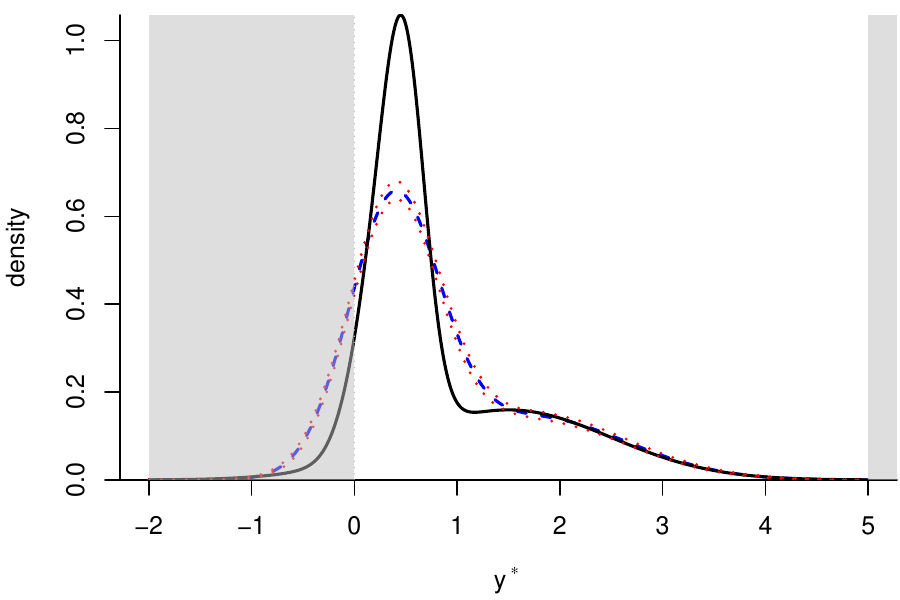}}
    \end{subfloat} \\ 
    \begin{subfloat}[ Three components \label{fig:app43}]
        {\includegraphics[width=0.49\textwidth]{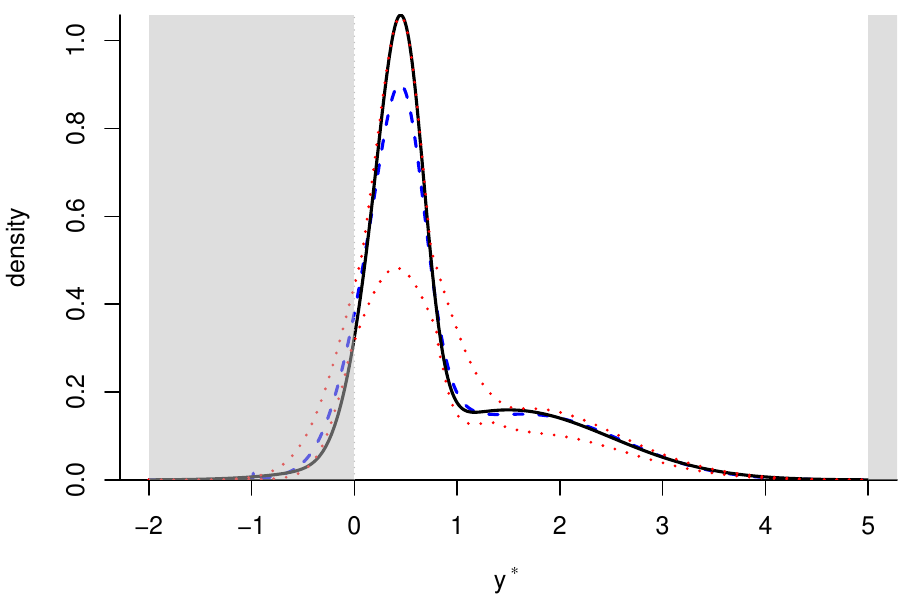}}
    \end{subfloat} 
    \begin{subfloat}[ Four components \label{fig:app44}]
        {\includegraphics[width=0.49\textwidth]{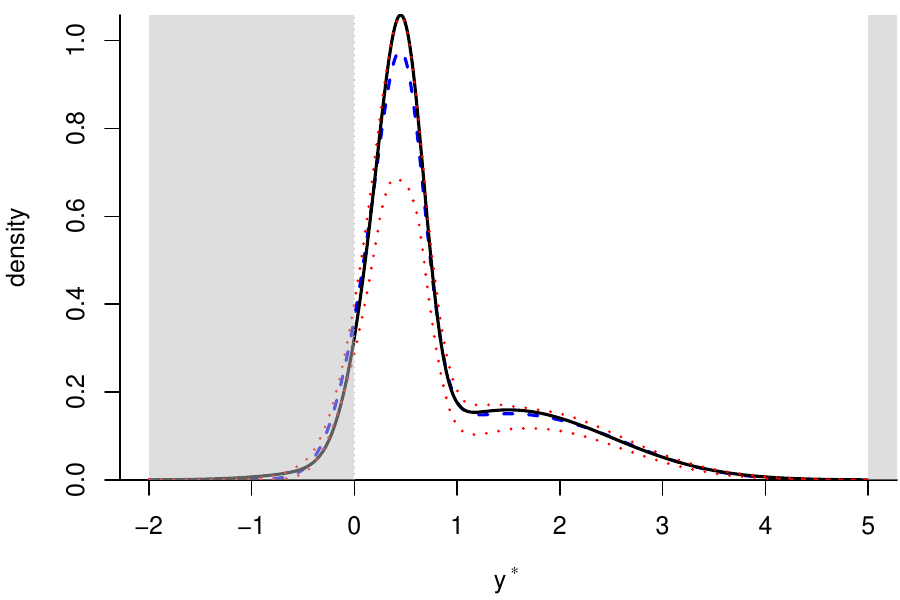}}
    \end{subfloat} 
\caption{Approximations to DGP 4}
    \label{fig:app4}%
\end{figure}

Finally, the approximations from applying the method proposed in Section \ref{sec:bayes} to data from DGP 4 are shown in Figure \ref{fig:app4}. As Figure \ref{fig:true_dists4} shows, this is the most challenging of the DGPs here considered for this model to handle: not only are the three components close together, but there is also visible censoring that affects all these components. Consequently, two of the components become visually indistinguishable and the remaining one becomes less pronounced.

Expectedly, the standard Tobit model cannot provide a good approximation to this model. In turn, the model with two components provides a more reasonable approximation. As with DGP 2, this is possibly a consequence of the resulting smoothness of  the mixture distribution.

As before, the models with three and four components perform best in approximating the underlying DGP. However, as with DGP 3, the latter provides a slightly better and more precisely estimated approximation than the former. Once again, this suggests that the flexibility of this method can be valuable given finite samples and unfavorable DGPs in terms of the extent of censoring.

\section{Applications}\label{sec:applies}

I consider three empirical applications meant to assess different aspects of the proposed methodology. First, I evaluate whether the model can yield substantially different estimates of the effects of a covariate of interest using data from a job training program experiment. Second, I show that the mixture model provides a better overall fit using data on labor supply, and that it further alters the conclusions about comparisons between annual hours worked of married and unmarried woman. Third, I investigate whether this new approach can handle dependent variables for which it might not be the theoretically appropriate model using data on demand for medical care. All data sets can be accessed via the \texttt{R} package \texttt{Ecdat}.

\subsection{A Job Training Program Experiment}\label{sec:lalonde}

I first utilize the data from the job training experiment considered by \cite{lalonde1986}. More specifically, I use the data from \cite{dw1999}---to which I direct the reader for a more detailed description---and compare Tobit models with different number of components and their implications for the measurement of the effect of job training programs. 

\subsubsection{Specification and Estimation}\label{sec:lalonde_spec}

I set real annual earnings in 1978 as the dependent variable and an indicator for whether the individual received the program as the covariate of interest. As in \cite{ct2005}, I also add age, age squared, education in years, real annual earnings in 1974 and 1975, and indicators for whether education was less than 12 years, the individual was black, and the individual was Hispanic as covariates. The minimum earnings are zero and 331 of the 2675 observations are censored (12.37\%).

For estimation, I obtained 100,000 draws, dropped the first half, and kept each 100th draw of the remaining draws. I used the maximum likelihood estimates of the standard Tobit model multiplied by randomly drawn numbers from a uniform distribution between 0.8 and 1.2 as the initial values for the $\beta$s and $\sigma$s and $\pi_c=1/C$ for all $c$. The parameters of the priors were $\alpha_c=1/C$, $\mu_c=0_{p\times 1}$, $\Omega_c=\tau\times I_{p\times p}$, where $\tau=0$, and $a_c=b_c=0$ for all $c$.

\subsubsection{Model Fit}

I only consider models with one and two components as I found that the mixing probability of an additional component would collapse to zero as the algorithm progressed. To assess which of the models fits the data better, Table \ref{tab:lalonde_ics} shows different metrics to assess and compare model fit: AIC, BIC, DIC, WAIC, and LOO.\footnote{I computed the WAIC and LOO using the \texttt{waic} and \texttt{loo} functions from the \texttt{R} package \texttt{loo}. While there were a few observations for which the estimated Pareto shape parameter exceeded the typical threshold, separate calculations, without Pareto smoothed importance sampling, yielded very similar results. Hence, I opted to continue using these functions to guarantee comparability of the results presented in Section \ref{sec:app_dt}.} All metrics suggest that the model with two components is preferable.



\begin{table}[H]
\begin{threeparttable}
\caption{Measures of Model Fit: \cite{lalonde1986}/\cite{dw1999} Data}
\label{tab:lalonde_ics}
\begin{tabular}{c|cc}
    \hline \hline 
\multirow{2}{*}{Criterion} & \multicolumn{2}{c}{\# of components}   \\ \cline{2-3}
 &  1 & 2    \\ 
\hline
AIC & 50,964.90 & \textbf{49,983.40} \\
BIC & 51,112.52 & \textbf{50,296.42} \\
DIC & 50,917.04 & \textbf{49,885.84} \\
WAIC & 50,972.77 & \textbf{49,992.57} \\
LOO & 50,972.67 & \textbf{49,992.82} \\
 \hline \hline
    \end{tabular}
\end{threeparttable}
  \end{table} 

\subsubsection{Estimates of the Effect of the Job Training Program}

\begin{figure}
    \centering
    \includegraphics[width=0.75\textwidth]{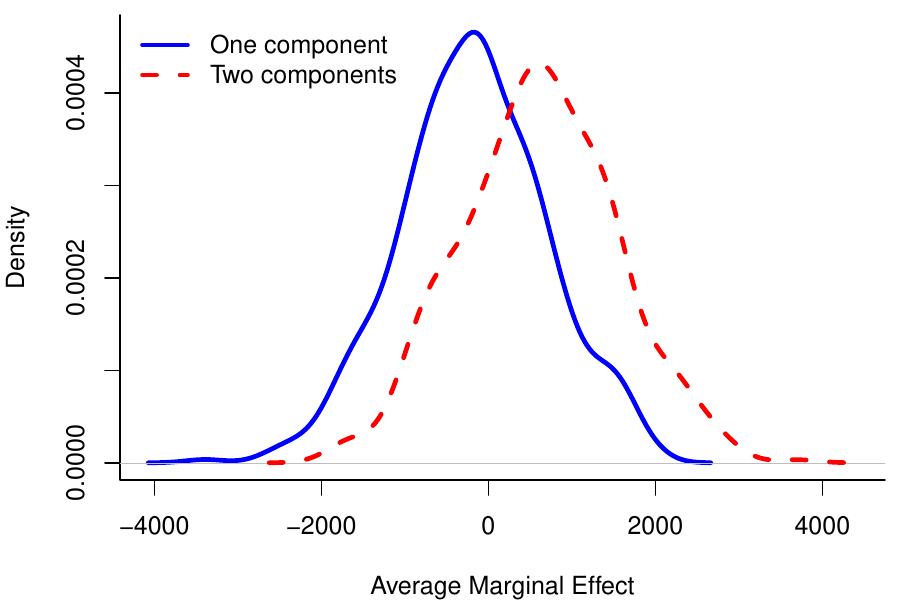}
    \caption{Average Marginal Effects of Job Training Program\label{fig:lalonde}}
\end{figure}

In the context of this job training program, arguably the researcher's main interest is not to obtain a good fit of the data but rather recover sensible estimates of the effect of the program on the workers, and in particular on their earnings. It is important to note that the current exercise is solely meant for illustration of the methodology and not to propose a replacement for the rigorous causal inference techniques motivated by \cite{lalonde1986} and \cite{dw1999}. To further assess the differences in results between the two models under consideration, I plot the density of the average marginal effects of the program across the draws for both models. Results are shown in Figure \ref{fig:lalonde}.

An additional component to the model slightly increases the dispersion of the estimates of the average marginal effect of the program as their standard deviation increases from 879.60 to 927.65. More importantly, the additional effectively shifts their distribution toward higher values, and the mean increases from -187.29 to 605.41.

This has an important consequence for the interpretation of the results. A simple comparison of the outcome variable between the treatment and control groups yields a negative estimate of the effect of program. Although including the covariates considered into a linear regression reverses this counterintuitive result,\footnote{The difference in outcomes is -15,505, whereas the OLS estimate of the coefficient associated with treatment becomes 218 after the inclusion of the covariates.} such model ignores the censoring of the dependent variable, and accounting for it using the standard Tobit model yields a negative estimate. 

However, providing more flexibility to the model by adding an additional component yields a positive estimate while accounting for the censoring of the outcome variable. This demonstrates that the flexibility of the proposed model can significantly impact estimates of the effects of covariates of interest.

\subsection{Labor Supply}

I now utilize data from the PSID to study the labor supply of women. In particular, I assess whether providing more flexibility to the standard Tobit model via additional components can improve the ability of the model to capture key elements of the distribution of annual work hours of women beyond improving overall model fit. I focus on differences between married and unmarried woman, which has received considerable attention in the literature.


\subsubsection{Specification and Estimation}\label{sec:psid_spec}

The outcome variable is annual work hours, which equals zero for 1,189 out of 4,855 observations (24.49\%).\footnote{One observation was dropped because the value of the education variable was missing.} I include age, education, number of children, and an indicator for whether the woman was married as covariates.

Estimation followed the guidelines from Section \ref{sec:lalonde_spec} except that in this case I set $\tau=0.0001$, obtained 150,000 draws, and kept each 300th draw of the second half of these draws. These modifications were required to ensure convergence and to mitigate the correlation between draws.

\subsubsection{Model Fit}

As in Section \ref{sec:lalonde}, I only consider models with one and two components as the mixing probability of a third component converged to zero. Table \ref{tab:psid_ics} is analogous to Table \ref{tab:lalonde_ics} and presents measures of fit between the models with one and two components. As before, all measures suggest that the model with two components is preferable.

\begin{table}[H]
\begin{threeparttable}
\caption{Measures of Model Fit: PSID Data}
\label{tab:psid_ics}
\begin{tabular}{c|cc}
    \hline \hline 
\multirow{2}{*}{Criterion} & \multicolumn{2}{c}{\# of components}   \\ \cline{2-3}
 &  1 & 2    \\ 
\hline
AIC & 64,845.75 & \textbf{64,663.37} \\
BIC & 64,931.60 & \textbf{64,854.05} \\
DIC & 64,819.47 & \textbf{64,612.30} \\
WAIC & 64,840.99 & \textbf{64,653.38} \\
LOO & 64,841.11 & \textbf{64,653.57} \\
 \hline \hline
    \end{tabular}
\end{threeparttable}
  \end{table} 

\subsubsection{Differences in Labor Supply of Married and Unmarried Women}

Even though the results from Table \ref{tab:psid_ics} suggest that the model with two components provides a better overall fit of the data, this need not imply that it yields substantial differences in key moments of the distribution of annual hours worked. Similarly to \cite{olson1998}, I focus on the probability of a married woman not working, her average annual hours worked conditional on working along with the unconditional average, and the differences between these objects and the analogous ones for unmarried women. I compute these objects using the averages over draws of the parameters and evaluate them at the sample averages of age, education, and number of children, and display the results in Table \ref{tab:psid_res}.

\begin{table}[H]
\begin{threeparttable}
\caption{Measures of Model Fit: PSID Data}
\label{tab:psid_res}
\begin{tabular}{c|cc}
    \hline \hline 
\multirow{2}{*}{Object} & \multicolumn{2}{c}{\# of components}   \\ \cline{2-3}
 &  1 & 2    \\ 
\hline
$\Pr \left ( \text{hours}=0 \middle \vert \text{married} \right )$ & 0.1812 & 0.2647 \\
$\Pr \left ( \text{hours}=0 \middle \vert \text{married} \right ) - \Pr \left ( \text{hours} \middle \vert \text{unmarried} \right )$ & -0.0193 & -0.0302 \\
$\mathbb{E} \left [ \text{hours} \middle \vert \text{hours}>0, \text{married} \right ]$ & 1,480.94 & 1,237.00 \\
$\mathbb{E} \left [ \text{hours} \middle \vert \text{hours}>0, \text{married} \right ] - \mathbb{E} \left [ \text{hours} \middle \vert \text{hours}>0, \text{unmarried} \right ]$ & 50.44 & 52.54 \\
$\mathbb{E} \left [ \text{hours} \middle \vert \text{married} \right ]$ & 1,212.60 & 909.56 \\
$\mathbb{E} \left [ \text{hours} \middle \vert \text{married} \right ] - \mathbb{E} \left [ \text{hours} \middle \vert \text{unmarried} \right ]$ & 68.84 & 74.41 \\
 \hline \hline
    \end{tabular}
\end{threeparttable}
  \end{table} 

The additional component alters the estimates in non-trivial ways. It considerably increases the probability that a married woman does not work from 18.12\% to 26.47\% while decreasing the average annual work hours from 1,480.94 to 1,237. These two decreases mechanically imply a reduction of unconditional labor supply, from 1,212.60 annual work hours to 909.56.

On the other hand, the differences between these objects and analogous ones for unmarried woman are intensified by the addition of a second component. Whereas the model with one component implies a difference in probabilities of not working of 1.93\%, the prediction under an additional component is higher, of 3.02\%. Furthermore, the differences in hours worked are higher under the model with components.

Put together, the results from Tables \ref{tab:psid_ics} and \ref{tab:psid_res} show that adding more flexibility to the model can not only increase its overall fit, but also alter specific quantities of interest as in Section \ref{sec:lalonde}. In this case, the main difference obtained from the added flexibility concerns the probability that a married woman works. In addition, the added flexibility can further change estimates of differences between groups, as illustrated by the comparison of labor supply of married and unmarried women.

\subsection{Demand for Medical Care}\label{sec:app_dt}

I now utilize the data set employed by \cite{dt1997}. The outcome of interest is the number of visits to a doctor, which is a count variable. Thus, this exercise is meant to investigate the performance of the proposed methodology in a setting where it might not be the most natural modeling approach.

\subsubsection{Specification and Estimation}

For simplicity, I only consider the number of physician office visits as the dependent variable. Out of the 4,406 observations, 683 (15.5\%) are censored. Following \cite{dt1997}, I include the following variables in the vector $x$: indicators for whether the patient's self-perceived health was excellent and poor, for whether they had a condition that limited activities of daily living, for whether they lived in northeastern, midwestern, or western US, for whether they were African American, male, married, and employed, and for whether they were covered by Medicaid or private health insurance, as well as their number of chronic conditions, age, family income, and years of schooling. Estimation followed the exact same procedure as described in Section \ref{sec:lalonde_spec}.

\subsubsection{Model Fit}

I consider models with one, two, and three components as the probability of an additional component collapsed to zero as the algorithm progressed. Table \ref{tab:dt_ics} shows the measures of model fit analogously to Tables \ref{tab:lalonde_ics} and \ref{tab:psid_ics}. All indicate that the model with three components is preferable.

\begin{table}[H]
\begin{threeparttable}
\caption{Measures of Model Fit: \cite{dt1997} Data}
\label{tab:dt_ics}
\begin{tabular}{c|ccc}
    \hline \hline 
\multirow{2}{*}{Criterion} & \multicolumn{3}{c}{\# of components}   \\ \cline{2-4}
 &  1 & 2 & 3    \\ 
\hline
AIC & 26,525.61 & 24,914.09 & \textbf{24,572.43} \\
BIC & 26,787.67 & 25,457.00 & \textbf{25,396.19} \\
DIC & 26,449.99 & 24,767.12 & \textbf{24,358.25} \\
WAIC & 26,530.46 & 24,916.78 & \textbf{24,564.57} \\
LOO & 26,530.79 & 24,916.68  & \textbf{24,568.12} \\
 \hline \hline
    \end{tabular}
\end{threeparttable}
  \end{table} 

\subsubsection{Comparison to Alternative Models}

\begin{figure}[H]
    \centering
    \includegraphics[width=0.75\textwidth]{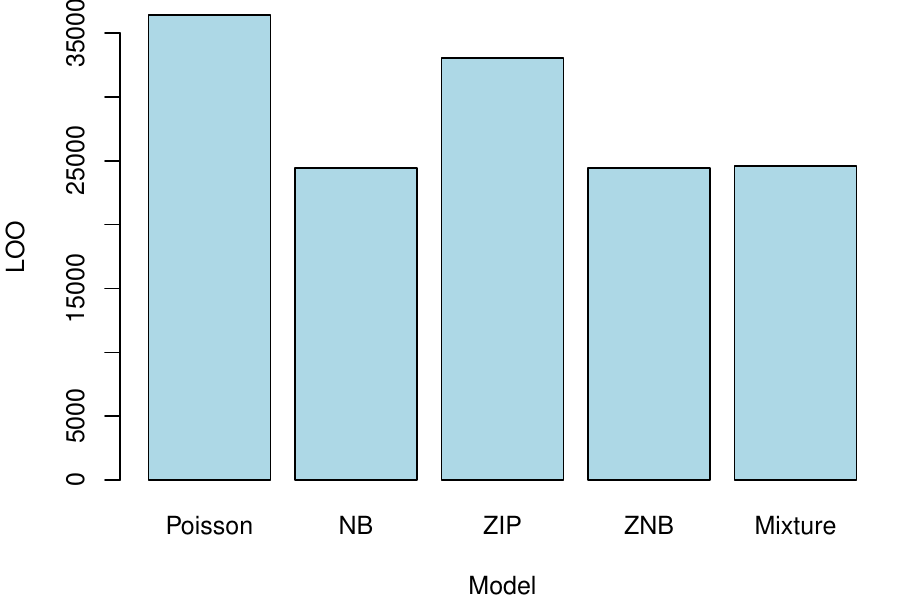}
    \caption{LOO of Different Models\label{fig:dt}}
\end{figure}

However, despite its flexibility a mixture of censored normal distributions arguably might not the most appropriate model in this case because the outcome of interest, number of doctor visits, is a count variable. To address whether the proposed model can be a good alternative in this scenario, I compare the mixture model with three components to four commonly used count data models: Poisson, zero-inflated Poisson (ZIP), negative binomial (NB), and zero-inflated negative binomial (ZNB).\footnote{These models were estimated using the \texttt{brm} function from the \texttt{R} package \texttt{brms}.} I perform this comparison using LOO as the criterion of model fit and present the results in Figure \ref{fig:dt}.

With a LOO equal to 24,568.12, as shown in Table \ref{tab:dt_ics}, the model with three components significantly outperforms the Poisson and ZIP models, whose LOOs equaled 36,414.92 and 33,066.79, respectively. In turn, the LOOS of the NB (24,446.51) and ZNB (24,446.94) models indicated that they slightly outperformed the mixture of three truncated normal distributions. 

Nevertheless, these results show that the proposed methodology can have a competitive and even significantly superior performance relative to models that are specifically tailored to handle count variables. This further showcases the versatility and general applicability of the finite mixture of Tobit models. 

A pertinent question is why this model should be considered in the context of a count outcome variable when other methods, which are arguably more natural, including the ones considered above, are available. When covariates are included in the model, such models typically require more involved techniques than the Gibbs sampling with data augmentation here proposed to obtain draws from the posterior distribution, such as Metropolis-Hastings and Hamiltonian Monte Carlo (e.g., \citealt{fla2023}). Hence, treating the count variable as continuous and using the finite mixture of Tobit models might be a more convenient alternative.

\section{Conclusion}\label{sec:conc}

This paper proposed a Bayesian approach to estimate finite mixtures of Tobit models. This approach combines techniques used to estimate finite mixtures of normal distributions and standard censored regression models and can be implemented through a simple Gibbs sampling algorithm with data augmentation, whose successful performance is confirmed via simulations.

In addition, I applied the proposed method to data on a job training program, labor supply, and demand for medical care. These applications show that the methodology can improve overall model fit, alter estimates of specific quantities of interest, such as the individual effect of specific variables, and perform well on data to which it might not be considered tailored to. These exercises showcase the flexibility and general applicability of the proposed methodology.


Despite its added flexibility, the model considered here still features significant structure. The conditional mean of the latent outcome variable on the covariates remains linear in parameters and there is no consideration of cases with high dimensionality. An interesting direction for future research could be relaxing these limitations, which might be achieved via the use of alternative prior distributions. In its current version, the choice of priors is motivated solely for convenience in order to ease and speed the implementation of the algorithm.

Furthermore, the number of components, $C$, is treated as fixed. In the empirical applications, I considered models with different values of $C$ and compared their ability to fit the data using popular metrics such as WAIC and LOO. A more rigorous approach would be to place priors on $C$ itself and estimate the posterior distribution along with the remaining parameters of the model. The methods considered by \cite{rg1997} and \cite{mh2018} are general enough to be applied with this model but require a more complex estimation approach.

Finally, another topic that I have not addressed is label switching. Under several circumstances the practitioner might be interested in the underlying classes themselves, in which case label switching can become a concern. More work is necessary to assess which techniques address this matter more effectively.

\newpage

\bibliography{mix_tobit}

\begin{thebibliography}{}

\bibitem[Amemiya, 1973]{amemiya1973}
Amemiya, T. (1973).
\newblock Regression analysis when the dependent variable is truncated normal.
\newblock {\em Econometrica}, 41(6):997--1016.

\bibitem[Arellano-Valle et~al., 2012]{avcgfmg2012}
Arellano-Valle, R.~B., Castro, L.~M., Gonz\'{a}lez-Far\'{i}as, G., and
  Mu\~{n}oz Gajardo, K.~A. (2012).
\newblock Student-$t$ censored regression model: Properties and inference.
\newblock {\em Statistical Methods \& Applications}, 21(4):453--473.

\bibitem[Busse et~al., 2017]{biz2017}
Busse, M.~R., Israeli, A., and Zettelmeyer, F. (2017).
\newblock Repairing the damage: The effect of price knowledge and gender on
  auto repair price quotes.
\newblock {\em Journal of Marketing Research}, 54(1):75--95.

\bibitem[Cameron and Trivedi, 2005]{ct2005}
Cameron, A.~C. and Trivedi, P.~K. (2005).
\newblock {\em Microeconometrics: Methods and Applications}.
\newblock Cambridge University Press.

\bibitem[Chib, 1992]{chib1992}
Chib, S. (1992).
\newblock Bayes inference in the {T}obit censored regression model.
\newblock {\em Journal of Econometrics}, 51(1--2):79--99.

\bibitem[Coughlan and Narasimhan, 1992]{cn1992}
Coughlan, A.~T. and Narasimhan, C. (1992).
\newblock An empirical analysis of sales-force compensation plans.
\newblock {\em Journal of Business}, 65(1):93--121.

\bibitem[Cui et~al., 2011]{czlm2011}
Cui, Y., Zhang, R., Li, W., and Mao, J. (2011).
\newblock Bid landscape forecasting in online ad exchange marketplace.
\newblock In Ghosh, J. and Smyth, P., editors, {\em Proc. of the 17th ACM
  SIGKDD Internat. Conf. on Knowledge Discovery \& Data Mining}, pages
  265--273. (ACM, New York).

\bibitem[Deb and Trivedi, 1997]{dt1997}
Deb, P. and Trivedi, P.~K. (1997).
\newblock Demand for medical care by the elderly: A finite mixture approach.
\newblock {\em Journal of Applied Econometrics}, 12(3):313--336.

\bibitem[Dehejia and Wahba, 1999]{dw1999}
Dehejia, R.~H. and Wahba, S. (1999).
\newblock Causal effects in nonexperimental studies: Reevaluating the
  evaluation of training programs.
\newblock {\em Journal of the American Statistical Association},
  448(94):1053--1062.

\bibitem[Diebolt and Robert, 1994]{dr1994}
Diebolt, J. and Robert, C.~P. (1994).
\newblock Estimation of finite mixture distributions through {B}ayesian
  sampling.
\newblock {\em Journal of the Royal Statistical Society, Series B},
  56(2):363--375.

\bibitem[Fagbamigbe et~al., 2023]{fla2023}
Fagbamigbe, A.~F., Lawal, T.~V., and Atoloye, K.~A. (2023).
\newblock Evaluating the performance of different {B}ayesian count models in
  modelling childhood vaccine uptake among children aged 12-23 months in
  {N}igeria.
\newblock {\em {BMC} Public Health}, 23(1197).

\bibitem[Garay et~al., 2015]{gblc2015}
Garay, A.~M., Bolfarine, H., Matos, L.~A., Lachos, V.~H., and Cabral, C. R.~B.
  (2015).
\newblock Bayesian analysis of censored linear regression models with scale
  mixture of normal distributions.
\newblock {\em Journal of Applied Statistics}, 42(12):2694--2714.

\bibitem[Horstmann and Moorthy, 2003]{hm2003}
Horstmann, I.~J. and Moorthy, S. (2003).
\newblock Advertising spending and quality for services: The role of capacity.
\newblock {\em Quantitative Marketing and Economics}, 1(3):337--365.

\bibitem[Hutton and Stanghellini, 2011]{hs2011}
Hutton, J.~L. and Stanghellini, E. (2011).
\newblock Modelling bounded health scores with censored skew-normal
  distributions.
\newblock {\em Statistics in Medicine}, 30(4):368--376.

\bibitem[Jedidi et~al., 1993]{jrd1993}
Jedidi, K., Ramaswamy, V., and Desarbo, W.~S. (1993).
\newblock A maximum likelihood method for latent class regression involving a
  censored dependent variable.
\newblock {\em Psychometrika}, 58(3):375--394.

\bibitem[Jindal, 2022]{jindal2022}
Jindal, P. (2022).
\newblock Perceived versus negotiated discounts: The role of advertised
  reference prices in price negotiations.
\newblock {\em Journal of Marketing Research}, 59(3):578--599.

\bibitem[Jo et~al., 2020]{jsct2020}
Jo, W., Sunder, S., Choi, J., and Trivedi, M. (2020).
\newblock Protecting consumers from themselves: Assessing consequences of usage
  restriction laws on online game usage and shopping.
\newblock {\em Marketing Science}, 39(1):117--133.

\bibitem[Keane and Stavrunova, 2011]{ks2011}
Keane, M. and Stavrunova, O. (2011).
\newblock A smooth mixture of {T}obits model for healthcare expenditure.
\newblock {\em Health Economics}, 20(9):1126--1153.

\bibitem[Kim and Kumar, 2018]{kk2018}
Kim, K.~H. and Kumar, V. (2018).
\newblock The relative influence of economic and relational direct marketing
  communications on buying behavior in business-to-business markets.
\newblock {\em Journal of Marketing Research}, 55(1):48--68.

\bibitem[Lachos et~al., 2019]{lcpd2019}
Lachos, V.~H., Cabral, C. R.~B., Prates, M.~O., and Dey, D.~K. (2019).
\newblock Flexible regression modeling for censored data based on mixtures of
  student-t distributions.
\newblock {\em Computational Statistics}, 34(1):123--152.

\bibitem[Lalonde, 1986]{lalonde1986}
Lalonde, R.~J. (1986).
\newblock Evaluating the econometric evaluations of training programs with
  experimental data.
\newblock {\em American Economic Review}, 76(4):604--620.

\bibitem[Lu et~al., 2020]{lmmr2020}
Lu, Y., Mitra, D., Musto, D., and Ray, S. (2020).
\newblock Can brands circumvent marketing regulations? {E}xploiting umbrella
  branding in financial markets.
\newblock {\em Marketing Science}, 39(1):71--91.

\bibitem[Massuia et~al., 2015]{mcml2015}
Massuia, M.~B., Cabral, C. R.~B., Matos, L.~A., and Lachos, V.~H. (2015).
\newblock Influence diagnostics for student-$t$ censored linear regression
  models.
\newblock {\em Statistics}, 49(5):1074--1094.

\bibitem[Massuia et~al., 2017]{mgcl2017}
Massuia, M.~B., Garay, A.~M., Cabral, C. R.~B., and Lachos, V.~H. (2017).
\newblock Bayesian analysis of censored linear regression models with scale
  mixtures of skew-normal distributions.
\newblock {\em Statistics and Its Interface}, 10(3):425--439.

\bibitem[Mattos et~al., 2018]{mgl2018}
Mattos, T. D.~B., Garay, A.~M., and Lachos, V.~H. (2018).
\newblock Likelihood-based inference for censored linear regression models with
  scale mixtures of skew-normal distributions.
\newblock {\em Journal of Applied Statistics}, 45(11):2039--2066.

\bibitem[Miller and Harrison, 2018]{mh2018}
Miller, J.~W. and Harrison, M.~T. (2018).
\newblock Mixture models with a prior on the number of components.
\newblock {\em Journal of the American Statistical Association},
  118(521):340--356.

\bibitem[Misra et~al., 2005]{mcn2005}
Misra, S., Coughlan, A.~T., and Narasimhan, C. (2005).
\newblock Salesforce compensation: An analytical and empirical examination of
  the agency theoretic approach.
\newblock {\em Quantitative Marketing and Economics}, 3(1):5--39.

\bibitem[Norets, 2010]{norets2010}
Norets, A. (2010).
\newblock Approximation of conditional densities by smooth mixtures of
  regressions.
\newblock {\em Annals of Statistics}, 38(3):1733--1766.

\bibitem[Olsen, 1978]{olsen1978}
Olsen, R.~J. (1978).
\newblock Note on the uniqueness of the maximum likelihood estimator for the
  {T}obit model.
\newblock {\em Econometrica}, 46(5):1211--1215.

\bibitem[Olson, 1998]{olson1998}
Olson, C.~A. (1998).
\newblock A comparison of parametric and semiparametric estimates of spousal
  health insurance coverage on weekly hours worked by wives.
\newblock {\em Journal of Applied Econometrics}, 13(5):543--565.

\bibitem[Richardson and Green, 1997]{rg1997}
Richardson, S. and Green, P.~J. (1997).
\newblock On {B}ayesian analysis of mixtures with an unknown number of
  components (with discussion).
\newblock {\em Journal of the Royal Statistical Society, Series B},
  59(4):731--792.

\bibitem[Rossi, 2014]{rossi2014}
Rossi, P.~E. (2014).
\newblock {\em Bayesian {N}on- and {S}emi-parametric {M}ethods and
  {A}pplications}.
\newblock Princeton University Press.

\bibitem[Tobin, 1958]{tobin1958}
Tobin, J. (1958).
\newblock Estimation of relationships for limited dependent variables.
\newblock {\em Econometrica}, 26(1):24--36.

\bibitem[Vana et~al., 2018]{vlb2018}
Vana, P., Lambrecht, A., and Bertini, M. (2018).
\newblock Cashback is cash forward: Delaying a discount to entice future
  spending.
\newblock {\em Journal of Marketing Research}, 55(6):852--868.

\bibitem[Zeller et~al., 2019]{zclb2019}
Zeller, C.~B., Cabral, C. R.~B., Lachos, V.~H., and Benites, L. (2019).
\newblock Finite mixture modeling of regression models for censored data based
  on scale mixtures of normal distributions.
\newblock {\em Advances in Data Analysis and Classification}, 13(1):89--116.

\end{thebibliography}

\end{document}